\pdfoutput=1

\documentclass[11pt,twoside,a4paper,cmspaper,final,collab]{cms-tdr}
\def\svnVersion{39604:40656M}\def\cmsCernNo{2010-093}\def\cmsCernDate{\today}\def\cmsMessage{Submitted to the Journal of High Energy Physics}

\begin{document}\cmsNoteHeader{BPH-10-010}

\hyphenation{had-ron-i-za-tion}
\hyphenation{cal-or-i-me-ter}
\hyphenation{de-vices}
\RCS$Revision: 39584 $
\RCS$HeadURL: svn+ssh://alverson@svn.cern.ch/reps/tdr2/papers/BPH-10-010/trunk/BPH-10-010.tex $
\RCS$Id: BPH-10-010.tex 39584 2011-02-15 18:37:39Z arizzi $
\newcommand{\bhadb}{{\PB\PaB\ }}
\newcommand{\bhadbbr}{{\PB\PaB\,}}
\newcommand{\bbar}{\ensuremath{{\overline{\mathrm{b}}}}\xspace}
\newcommand{\cbar}{\ensuremath{{\overline{\mathrm{c}}}}\xspace}
\newcommand{\CASCADE}{{\textsc{cascade}}\xspace}
\newcommand{\Pbquark}{\ensuremath{\mathrm{b}}\xspace}
\newcommand{\Pbbbar}{\ensuremath{\mathrm{b\overline{b}}}\xspace}

\cmsNoteHeader{BPH-10-010} 
\title{Measurement of \PB\PaB\ Angular Correlations based on Secondary Vertex
Reconstruction at $\sqrt{s}=7$ TeV}

\address[neu]{ETH Zurich}
\address[fnal]{Fermilab}
\address[cern]{CERN}
\author[cern]{The CMS Collaboration}

\date{\today}

\abstract{
A measurement of the angular correlations between beauty and anti-beauty hadrons (\PB\PaB)
produced in pp collisions at a centre-of-mass energy of $ 7\; \mbox{TeV}$
at the CERN LHC is presented, probing for the first time the region of small angular separation.
The \PB\ hadrons are identified by the presence of displaced
secondary vertices from their decays.
The \PB\ hadron angular separation is reconstructed from the decay
vertices and the primary-interaction vertex.
The differential \bhadb production cross section,
measured from a data sample collected by CMS and corresponding to an integrated luminosity of
$3.1\, \mathrm{pb}^{-1}$,
shows that a sizable fraction of the \bhadbbr pairs are produced
with small opening angles.
These studies provide a test of QCD and further insight into the dynamics of
\Pbbbar production.
}

\hypersetup{%
pdfauthor={CMS Collaboration},%
pdftitle={Measurements of B anti-B Angular Correlations based on Secondary Vertex
Reconstruction at sqrt_s=7 TeV},%
pdfsubject={CMS},%
pdfkeywords={CMS, physics, software, computing}}

\maketitle 

\setcounter{page}{2}%
\section{Introduction}

\label{s:introduction}
Beauty quarks are abundantly produced through strong interactions in pp collisions
at the CERN Large Hadron Collider (LHC). The hadroproduction of \Pbbbar pairs is measured
to have a large cross section (of the order of $100\,\mu{\rm b}$) 
at a centre-of-mass energy 
of $7\; \mbox{TeV}$~\cite{CMS:2011hf,ref:BPH-10-009,Aaij:2010gn}. 
Detailed \Pbquark quark production studies provide substantial information about the
dynamics of the underlying hard scattering subprocesses
within perturbative Quantum Chromodynamics (pQCD).
In lowest order pQCD, i.e.\ in $2\rightarrow 2$ parton interaction subprocesses,
momentum conservation requires the \Pbquark and \bbar quarks to be emitted in a
back-to-back topology.
However, higher order $2\rightarrow 2+n$ ($n\ge 1$) subprocesses with additional partons
(notably gluons) emitted,
give rise to different topologies of the final state \Pbquark quarks.
Consequently, measurements of \Pbbbar angular and momentum
correlations provide information about the underlying production subprocesses and allow for a
sensitive test of pQCD leading-order (LO) and next-to-leading order (NLO)
cross sections and their evolution with event energy scales.
Studies of b quark production at the LHC may provide
insight into the hadronisation properties of heavy quarks at these new energy scales,
as well as better knowledge of the heavy quark content of the proton.
In addition, identification of \Pbquark quarks and precision measurements of
their properties are crucial ingredients for new physics searches in which
\Pbbbar hadroproduction is expected to be one of the main backgrounds.

In this paper, angular correlations between pairs of beauty hadrons,
hereafter referred to as ``\PB\ hadrons'',
are studied with the Compact Muon Solenoid (CMS) detector,
probing for the first time the region of very small angular separation
at $\sqrt{s} = 7\; \mbox{TeV}$.
Measurements of \PB\PaB-pair production are presented differentially
as a function of the opening angle for different event scales,
characterised by the leading jet transverse momentum.
The extrapolation back to the angular separation of the  \Pbquark quarks,
which requires modeling of heavy quark fragmentation and hadronisation,
is not considered in this analysis. The results are given for
the visible kinematic range defined by the phase space at the
hadron level.

Measurements of the full range of \bhadb angular
separation demand good angular resolution and require the ability to resolve small
opening angles when the two \PB\  hadrons are inside a single reconstructed jet.
The kinematic properties of \PB\  hadrons can be reconstructed  using jets,
leptons from semileptonic decays of \PB\  hadrons
or secondary vertices (SV) originating from the decay of long-lived \PB\  hadrons.
In this analysis, a method based on an iterative inclusive secondary vertex finder
that exploits the excellent tracking capabilities of the CMS detector is introduced.
One advantage of this method is the unique capability
to detect  \bhadbbr pairs even at small opening angles,
in which case the decay products of the
\PB\  hadrons tend to be merged into a single jet and the
standard \PB\  jet tagging techniques~\cite{ref:BTV-10-001} are  not applicable.
Previously, studies of azimuthal \Pbbbar correlations using vertexing
have been done at lower energy
in \Pp\Pap\ collisions~\cite{Acosta:2004nj,Aaltonen:2007zza}.

In Section 2, a brief overview of the subdetectors relevant for this analysis
is given.
Section 3 describes the Monte Carlo (MC) simulations and the programs
used for QCD predictions.
The event selection, the analysis details,
and the determination of efficiencies and systematic uncertainties are
described in Section 4.
In Section 5 we present the results and compare the data with
theoretical predictions.

\section{The CMS Detector}
\label{s:detector}

A detailed description of the CMS detector can be found in Ref.~\cite{cms:2008zzk}.
The central feature of the CMS apparatus is a superconducting solenoid of 6 m internal
diameter, with a 3.8 T axial magnetic field.
The subdetectors used in the present analysis are tracking detectors and calorimeters,
located within the field volume.
The tracker consists of a silicon pixel and silicon strip tracker
covering the pseudorapidity range $|\eta| < 2.5$.
The pixel tracker consists of three barrel layers and two endcap disks at each barrel end.
The strip tracker has 10 barrel layers and 12 endcap disks.
The barrel and endcap calorimeters ($|\eta| < 3$) consist of a lead-tungstate
crystal electromagnetic calorimeter (ECAL)
and a brass/scintillator hadron calorimeter (HCAL).
The ECAL and HCAL cells are
grouped into towers, projecting radially outward from the interaction region,
for triggering purposes and to facilitate jet reconstruction.
The CMS experiment uses a right-handed coordinate system, with the origin at the nominal
proton-proton collision point, the $x$-axis pointing towards the centre of the LHC ring, the
$y$-axis pointing upwards (perpendicular to the LHC plane), and the $z$-axis pointing along the
anticlockwise beam direction. The polar angle $\theta$ is measured from the positive
$z$-axis and the azimuthal angle $\phi$ is measured from the positive $x$-axis in the $xy$ plane.
The radius $r$
denotes the distance from the $z$-axis and the pseudorapidity is defined by
$\eta = -\ln ( \tan(\theta / 2))$.

\section{Monte Carlo Simulation and QCD Predictions}
\label{s:mcsim}

Different simulation programs at the LO and the NLO level have been utilized to describe the
\Pbquark production process within perturbative QCD.
Within the LO picture, three parton level production subprocesses can be
defined \cite{Zanderighi:2007dy,Banfi:2007gu}, conventionally denoted by flavour creation (FCR),
flavour excitation (FEX) and gluon splitting (GSP),
and are implemented in Monte Carlo
event generators like {\PYTHIA}~\cite{Sjostrand:2006za} and
\HERWIG~\cite{Corcella:2000bw}.
These subprocesses are related to different final state topologies.
Notably, in FCR processes the \Pbbbar pairs are expected to be emitted in a
back-to-back topology, which corresponds to a large angular separation
between the \Pbquark and \bbar quarks, whereas in GSP the pair emission follows
a more collinear topology, i.e.\ a small angular separation between
the \Pbquark and \bbar quarks.
At higher orders in QCD, the FCR, FEX and GSP separation
of production subprocesses becomes meaningless and only the combination of the
$2\rightarrow 2$ and  $2\rightarrow 2+n$ ($n\ge 1$) subprocesses is relevant.
Calculations of such processes are implemented in
{\MCATNLO}~\cite{Frixione:2002ik, Frixione:2003ei, Frixione:2008ym} or
{\textsc{fonll}}~\cite{Cacciari:2008zb}.
The \MADGRAPH/{\textsc{madevent}}~\cite{Maltoni:2002qb,Alwall:2007st} generator
provides the possibility to simulate $2\rightarrow 2,3$ subprocesses at tree-level,
providing a hybrid solution between $2\rightarrow 2$ at LO and the NLO simulations.
We use also the {\textsc{cascade}}~\cite{Jung:2000hk} generator, which
is based on off-shell LO matrix
elements using high-energy factorization~\cite{Catani:1990eg} convolved
with unintegrated parton distributions.

The basic Monte Carlo event generator applied in this analysis is the LO {\PYTHIA}
program (version 6.422~\cite{Sjostrand:2006za}),
which is used to determine selection efficiencies and to optimise the vertexing
algorithm for \PB\  hadron reconstruction.
The event samples are generated applying the standard {\PYTHIA} settings~\cite{Sjostrand:2006za}
with {tune D6T}~\cite{Field:2010su} for the underlying event and with the
{CTEQ6L1}~\cite{Pumplin:2002vw} proton  parton distribution functions (PDF).
All events generated by the {\PYTHIA} program are processed with a detailed simulation
of the CMS detector response based on the {\textsc{geant4}} package~\cite{Agostinelli:2002hh}.

For comparison with theoretical predictions, events with two and three partons
in the final state are generated by means of the
 \MADGRAPH/{\textsc{madevent4}} program, where the
showering is performed with {\PYTHIA}, and the jet matching scheme
used is ``$k_\mathrm{T}$-MLM''~\cite{Alwall:2008qv}.
The {CTEQ6L1}~\cite{Pumplin:2002vw} parton distribution functions are used,
and the mass of the \Pbquark quark is set to m$_b = 4.75$~GeV.

For the events produced with the {\CASCADE} generator,
the CCFM set A~\cite{Jung:2010si} of parton distributions is used.
The calculations include the processes $g^* g^* \to b\bar{b}$ and $g^*q \to g q \to b\bar{b} X$.
The matrix element of $g^* g^* \to b\bar{b}$ already includes a large fraction
of the process $g^*g \to gg \to b\bar{b} X$~\cite{Catani:1990eg,Deak:2009xt},
therefore $g^*g \to gg \to b\bar{b} X$ is not added to avoid double counting.

A further set of QCD events is produced by means of the {\MCATNLO} generator
(version 3.4~\cite{Frixione:2008ym} with standard scale settings and
\Pbquark-quark mass m$_b = 4.75$~GeV),
which matches NLO QCD matrix element calculations with parton shower simulations as implemented
in {\HERWIG} (version 6.510)~\cite{Corcella:2000bw}.
The proton PDF set used is {CTEQ6M}~\cite{Pumplin:2002vw}.
For the NLO generated events, no full CMS detector simulation is done.
Subsequent to the parton showering and hadronisation process,
the generated stable particles in the events
are clustered into jets with the anti-$k_\mathrm{T}$ jet algorithm~\cite{Cacciari:2008gp}.

\section{Event Selection and Data Analysis}
\label{s:selection}

The data sample used in this analysis was collected by
the CMS experiment during 2010 at a
centre-of-mass energy of $\sqrt{s} = 7\; \mbox{TeV}$
and corresponds to an integrated luminosity of $3.1\,\pm\, 0.3\, \mathrm{pb}^{-1}$.
Only data from runs when the CMS detector components
relevant for this analysis were fully functional
and when stable beam conditions were present are used.
Events from non-collision processes are rejected by requiring
a  primary (``collision'') vertex (PV)~\cite{CMS-PAS-TRK-10-005, Khachatryan:2010pw}
 with at least four well reconstructed tracks.
Background from beam-wall and beam halo events, and events faking high energy deposits
in the HCAL, are filtered out
based on pulse shape, hit multiplicity and timing criteria.

\subsection{Analysis Overview}
\label{s:analysis}

\begin{figure}[htbp]
	\centering
		\includegraphics[angle=90,width=0.45\textwidth]{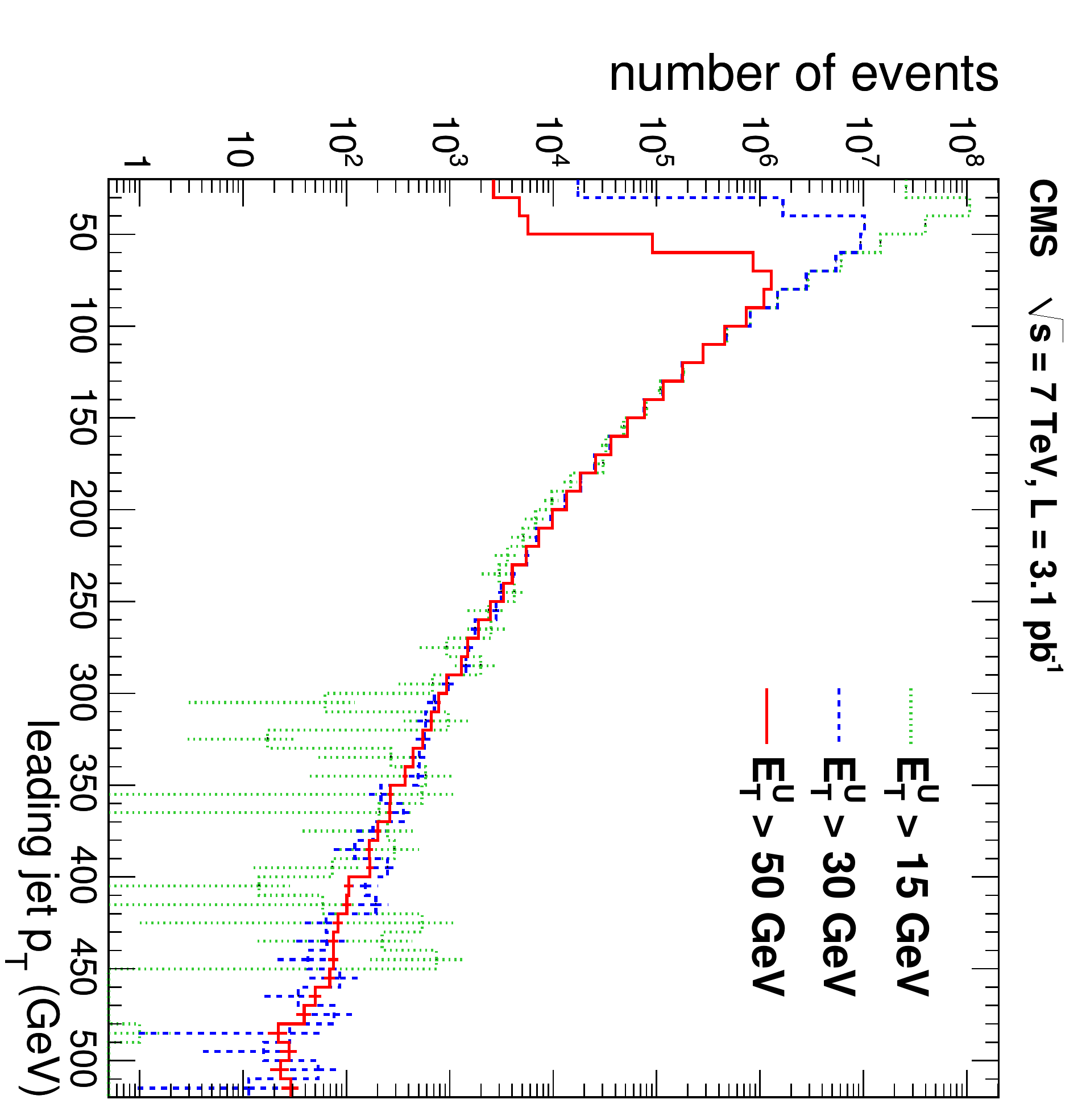}
		\includegraphics[angle=90,width=0.45\textwidth]{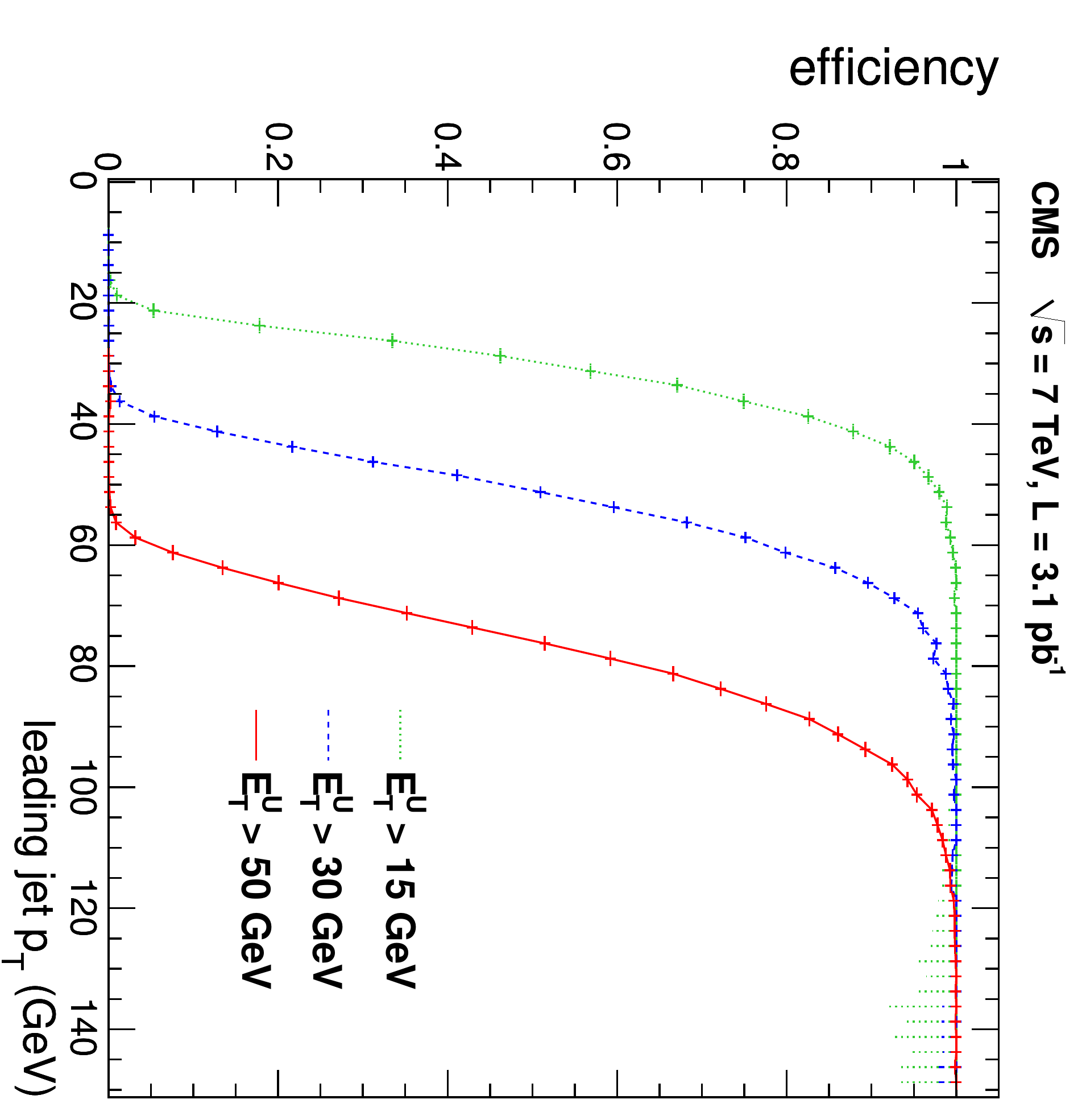}
	\caption{The measured transverse momentum distributions of the leading jet in the event (left) and
   measured efficiency to trigger an event on the high-level trigger
   as a function of  jet \pt\ (right), for three different trigger thresholds.
 }
\label{fig:hlt-turnon}
\end{figure}

The analysis relies on the single-jet trigger in both
the hardware-level (L1) and the software high-level (HLT) components
of the CMS trigger system~\cite{cms:2008zzk}.
We require at least one HLT jet with uncorrected transverse
calorimetric energy $E^U_\mathrm{T}$ above a trigger threshold of 15, 30 or 50 GeV.
Figure~\ref{fig:hlt-turnon} shows the
leading jet transverse momentum (\pt) spectra with particle
flow jets~\cite{CMS-PAS-PFT-10-002}
and the corresponding trigger efficiency dependence on \pt.
The efficiencies, also shown in Figure~\ref{fig:hlt-turnon}, are determined using events selected with a lower $E^U_\mathrm{T}$
(prescaled) trigger.

The event sample is then divided into three energy scale
bins corresponding to the \pt ranges
where the different jet triggers are over 99\% efficient.
These correspond to samples where the transverse momenta of the leading jet,
using corrected jet energies~\cite{CMS-PAS-JME-10-010},
exceed 56, 84 and 120~GeV, respectively.
The effective integrated luminosity, taking into account the trigger prescale factors,
corresponds to
$0.031$,
$0.313$ and
$3.069\, \mathrm{pb}^{-1}$, respectively,
for the three samples, including some overlap.

The visible kinematic range for the measurements is defined at the
\PB\  hadron level by the requirements
$|\eta(\mathrm{B})| < 2.0$ and $\pt(\mathrm{B}) > 15$ GeV for both of the \PB\  hadrons.
The leading jet used to define the minimum energy scale
is required to be within $|\eta(\mathrm{jet})| < 3.0$.

In this analysis, the HLT triggered events are required to have
at least one reconstructed jet with a minimum corrected \pt,
a reconstructed PV,
and in addition at least two
reconstructed secondary vertices (SV).
For the offline jet reconstruction, particle flow objects~\cite{CMS-PAS-PFT-10-002}
are clustered with the anti-$k_\mathrm{T}$ jet
algorithm~\cite{CMS-PAS-JME-10-003,Cacciari:2008gp}
with a distance parameter $R_{k_\mathrm{T}}=0.5$.
For further \bhadb angular analysis,
these generic secondary vertices are required to originate from  \PB\ hadron decays,
as described in the following paragraphs.

The flight direction of the original \PB\  hadron  is approximated by the
vector ${\overrightarrow {SV}}$, joining the PV (position of \PB\  hadron production)
and the SV (position of the \PB\  hadron decay).
The length $|{\overrightarrow {SV}}|$ is the three-dimensional  flight distance
($D_{3D}$) and its significance is given by
$S_{3D}=D_{3D}/ \sigma (D_{3D}) $, where $\sigma (D_{3D}) $ is
the uncertainty of $D_{3D}$.

In an event with two SVs, which are considered to originate from a \Pbbbar pair,
the angular correlation variables between
the \PB\  and \PaB\ hadrons are calculated using their flight directions.
Typical variables used for the characterization of the angular correlations
between the two hadrons are the difference in azimuthal angles ($\Delta \phi$)
and the difference in polar angles,
usually expressed in terms of pseudorapidity ($\Delta \eta$), or the combined separation
variable $\Delta R = \sqrt{\Delta \eta^2 + \Delta \phi^2}$.

The kinematic regions with $\Delta R < 0.8$ and
with $\Delta R >2.4$ are used for comparisons or normalisations of the
simulation.
The cross sections integrated over these two regions will be denoted
by $\sigma_{\Delta R < 0.8}$ and by $\sigma_{\Delta R >2.4}$,
and the ratio by $\rho_{\Delta R} = \sigma_{\Delta R < 0.8}/ \sigma_{\Delta R >2.4}$.
This is inspired by the theoretical predictions,
since at low $\Delta R$ values
the gluon splitting process is expected to contribute significantly,
whereas at high $\Delta R$ values flavour creation prevails.

\subsection{Vertex Reconstruction and \PB\  Candidate Identification}
\label{s:vertex}

The primary vertex is reconstructed from
tracks of low impact parameter with respect to the nominal
interaction region.
In cases of multiple interactions in the same bunch crossing (pile-up events),
the primary interaction vertex is chosen to be the one with
the largest squared transverse momentum sum $S_\mathrm{T} = \sum{{p^2_\mathrm{T}}_{i}}$,
where the sum runs over all tracks associated with the vertex.
Residual effects from pile-up events are found to be negligible.

Next, the events are required to have at least two reconstructed secondary vertices.
An inclusive secondary vertex finding (IVF) technique,
completely independent of jet reconstruction,
is applied for this purpose.
This technique reconstructs secondary vertices  by clustering tracks around the
so-called seeding tracks characterized  by high three-dimensional impact parameter
significance $S_d = d / \sigma(d)$, where $d$ and
$\sigma(d)$ are the impact parameter and its uncertainty at the PV, respectively.
The tracks are clustered to a seed track based on their
compatibility given their separation distance in three dimensions,
the separation distance significance (distance normalised to its uncertainty),
and the angular separation.
The clustered tracks are then
fitted to a common vertex with an outlier-resistant
fitter~\cite{Fruhwirth:2007hz,ref:AVR}.
The vertices sharing more than 70\% of the tracks
compatible within the uncertainties are merged. As a final step, all
tracks are assigned to either the primary or the secondary vertices on
the basis of the significance of the track to vertex distance.

In this analysis, a SV is required to be made up of at least three tracks,
to have a maximal two-dimensional flight distance
$D_{xy}=|{\overrightarrow {SV}}_{xy}| < 2.5$ cm,
a minimal  two-dimensional flight distance significance
 $S_{2D}=D_{xy}/ \sigma (D_{xy}) >3$, and
to possess a vertex mass $m_{SV} < 6.5$ GeV.
Here, $\sigma(D_{xy})$ is the uncertainty on $D_{xy}$.
The four-momentum of the vertex $p_{SV}= (E_{SV},{\vec p_{SV}} )$
is calculated as the sum $p_{SV} = \sum p_i$
over all tracks fitted to that vertex, with
$p_i =( E_i , {\vec p}_{i} )$, using the pion mass hypothesis
for every track to obtain its energy $E_i$.
The vertex mass $m_{SV}$ is calculated as
$m_{SV}^2 = E_{SV}^2 - {\vec p_{SV}}^2$.
The four-momentum of the reconstructed \PB\  hadron candidate is then
identified with the SV four--momentum, and thus
the variables $\pt(B), \eta(B)$ for the \PB\  hadron
candidates are readily calculated from $ p_{SV}$.

Events with at least two secondary vertices may
originate from any of the following processes:
a) true 'signal' \bhadb events;
b) true \bhadb events where at least one \PB\  hadron is not correctly reconstructed
(SV from other sources);
c) QCD events with light quark and gluon jets, which enter through misidentification
of vertices not originating from \PB\ decay;
d) direct \ccbar production with long lived \PD\ hadrons;
e) sequential $B \rightarrow D \rightarrow X$ decay chains, where
  \PB\  hadrons decay to long lived \PD\ hadrons, and
  both \PB\  and \PD\ vertices are reconstructed.
The \bhadb signal events contain a fraction from top quark pair production
of less than 1\%~\cite{Khachatryan:2010ez,Aad:2010ey}.

Often, both the \PB\  and \PD\ decay vertices are reconstructed by the IVF.
Such topologies need to be distinguished from
events with two quasi-collinear \PB\  hadrons.
To achieve this, an iterative merging procedure is applied to vertices with $\Delta R < 0.4$.
The procedure is optimised to yield a single \PB\  candidate associated
with a decay chain $B \rightarrow D \rightarrow X$,
while successfully retaining two \PB\  candidates also in events where
two real \PB\  hadrons are emitted nearly collinearly.
The vertices are merged into a single \PB\  candidate
if the invariant mass of the sum over all tracks is below 5.5 GeV and
 $\cos \beta > 0.99$, where $\beta$ is the angle
between the line connecting the two vertices and the sum of the momenta of the tracks
associated to the vertex at largest distance from the PV.

\begin{figure}[htbp]
	\centering
       \includegraphics[angle=90,width=0.45\textwidth]{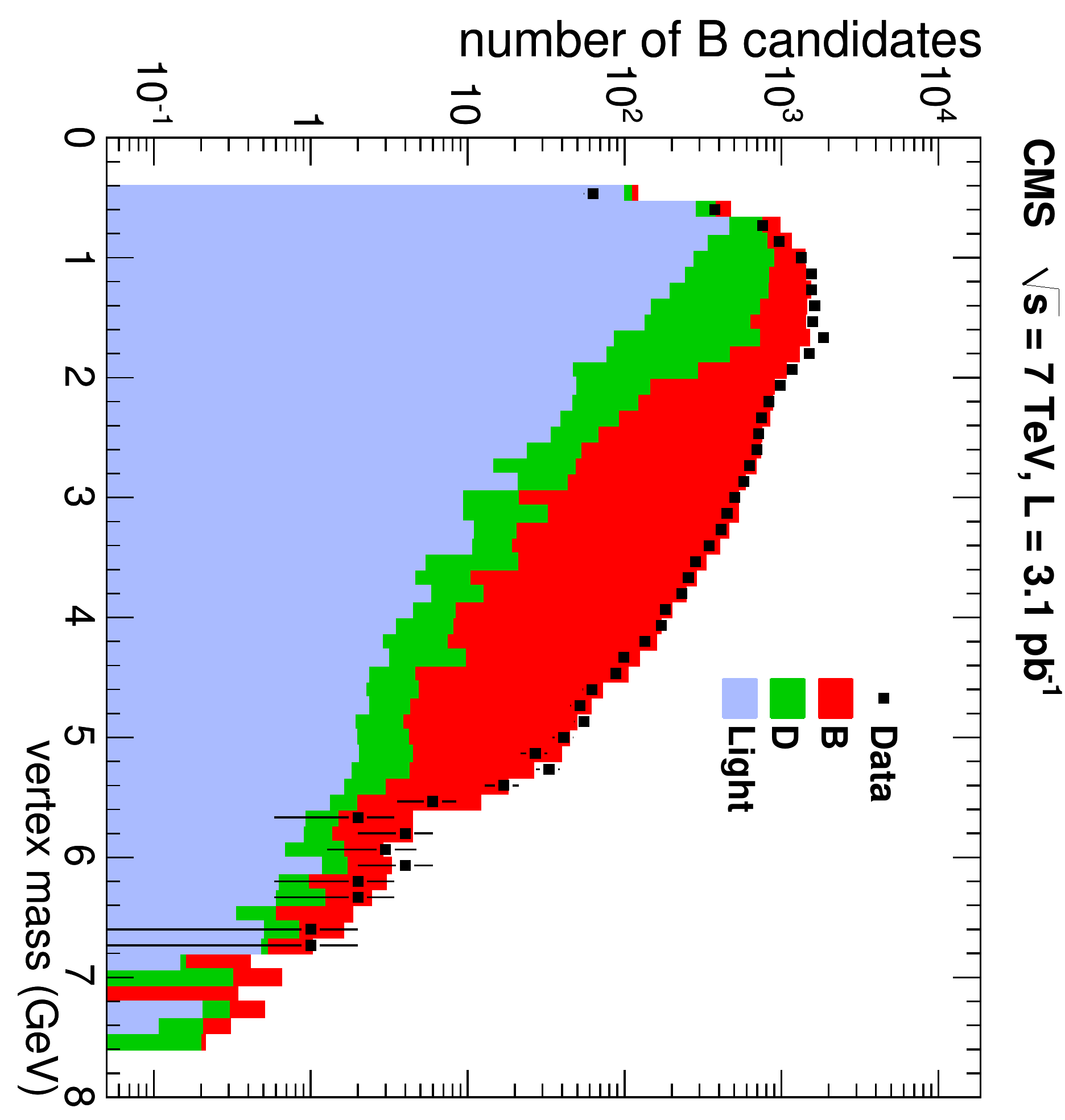}
        \includegraphics[angle=90,width=0.45\textwidth]{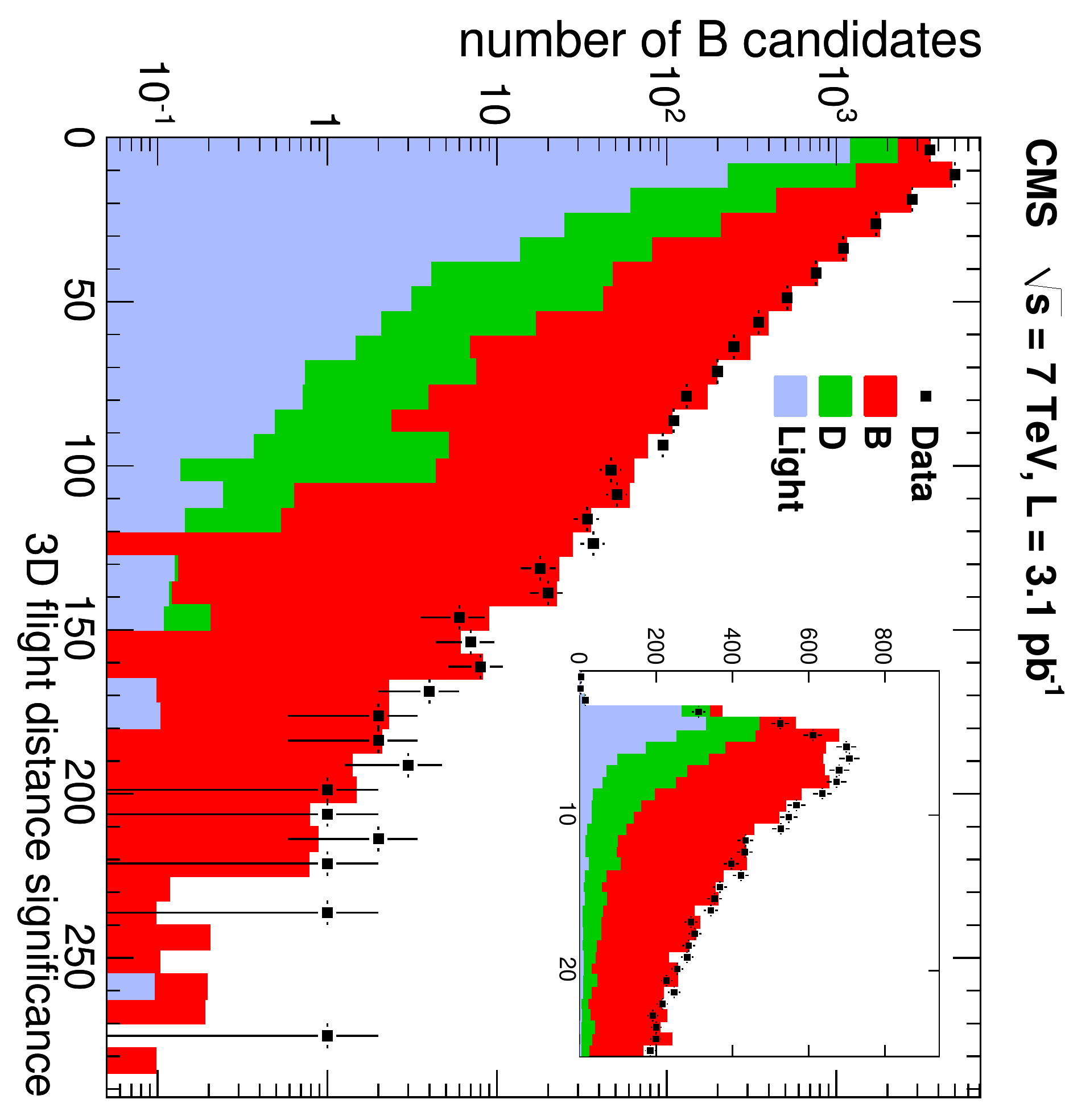}
	\caption{Properties of the reconstructed \PB\  candidates:
    vertex mass distribution (left) and flight distance significance distribution (right). The inset in the right plot shows a zoom of the flight distance significance distribution with narrower bins and linear scale.
     The data are shown by the solid points.
   The decomposition into the different sources, beauty, charm and light quarks,
   is shown for the {\PYTHIA} Monte Carlo simulation.
   The simulated distributions are normalised to the total number of data events.
   All selection cuts apart from those on the shown quantities are applied.
   }
\label{fig:SVproperties}
\end{figure}

All \PB\  candidates are retained if they
have a minimal 3D flight distance significance
$S_{3D} > 5$, a pseudorapidity $|\eta(\mathrm{SV})| < 2$,  a transverse momentum $\pt(\mathrm{SV}) > 8$ GeV,
and a vertex mass $m_{\mathrm{SV}} > 1.4$~GeV.
The quality of the \PB\  candidate reconstruction technique
is illustrated in Fig.~\ref{fig:SVproperties}
for events with a leading jet having $\pt > 84$ GeV
(all selection cuts apart from those on the shown quantities are applied).
The simulation describes the data very well in terms of vertex mass and 3D
decay length significance distribution.

Only those events which have exactly two \PB\  hadron candidates
and which have a vertex mass sum $m_1 + m_2 > 4.5$~GeV are retained.
A total of 160, 380 and 1038 events pass all these requirements for the
three leading jet \pt bins, respectively from the lowest to the highest.
The overall contributions from events with three or more \PB\  candidates
is found to be negligible (less than 1\%).

\subsection{Efficiency and Resolution}
\label{s:efficiency}

This analysis uses selection efficiency corrections as a function of
the leading jet \pt and the $\Delta R$ between the two SVs.
The corrections are determined from the simulated {\PYTHIA} event samples.
They extrapolate from the measured vertex momenta to the
visible phase space of true \PB\ hadrons, defined by
$|\eta(\mathrm{B})| < 2.0$, and $\pt({\mathrm{B}}) > 15$ GeV.
The momentum measured by the vertex candidate
represents of the order of 50\% of the true \PB\  hadron momentum.
The overall event reconstruction efficiencies (including both \PB\ hadron
decays) are found to be 7.4\%, 9.3\% and 10.7\%, on average,
for the three jet \pt bins, respectively from the lowest to the highest.

The validity of the $\Delta R$-dependence of the efficiencies
obtained from simulation is checked using a data driven method based on
event mixing, as illustrated below.
It is found that the $\Delta R$-dependence is well described  by the simulation,
justifying this approach. The differences are used to estimate the
systematic uncertainties.

\begin{figure}[htb]
\centering
\includegraphics[angle=90,width=0.45\textwidth]{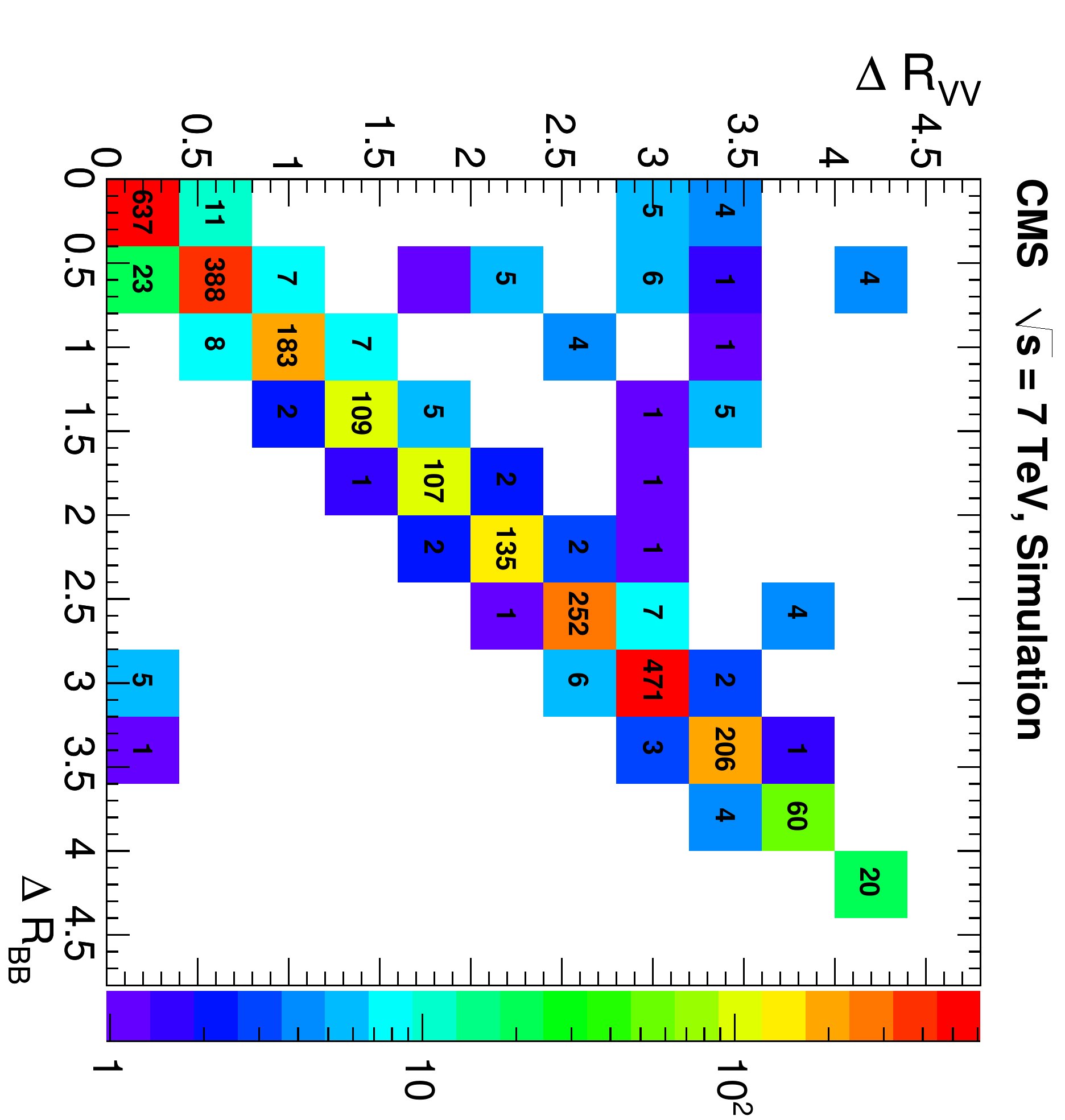}
\includegraphics[angle=90,width=0.45\textwidth]{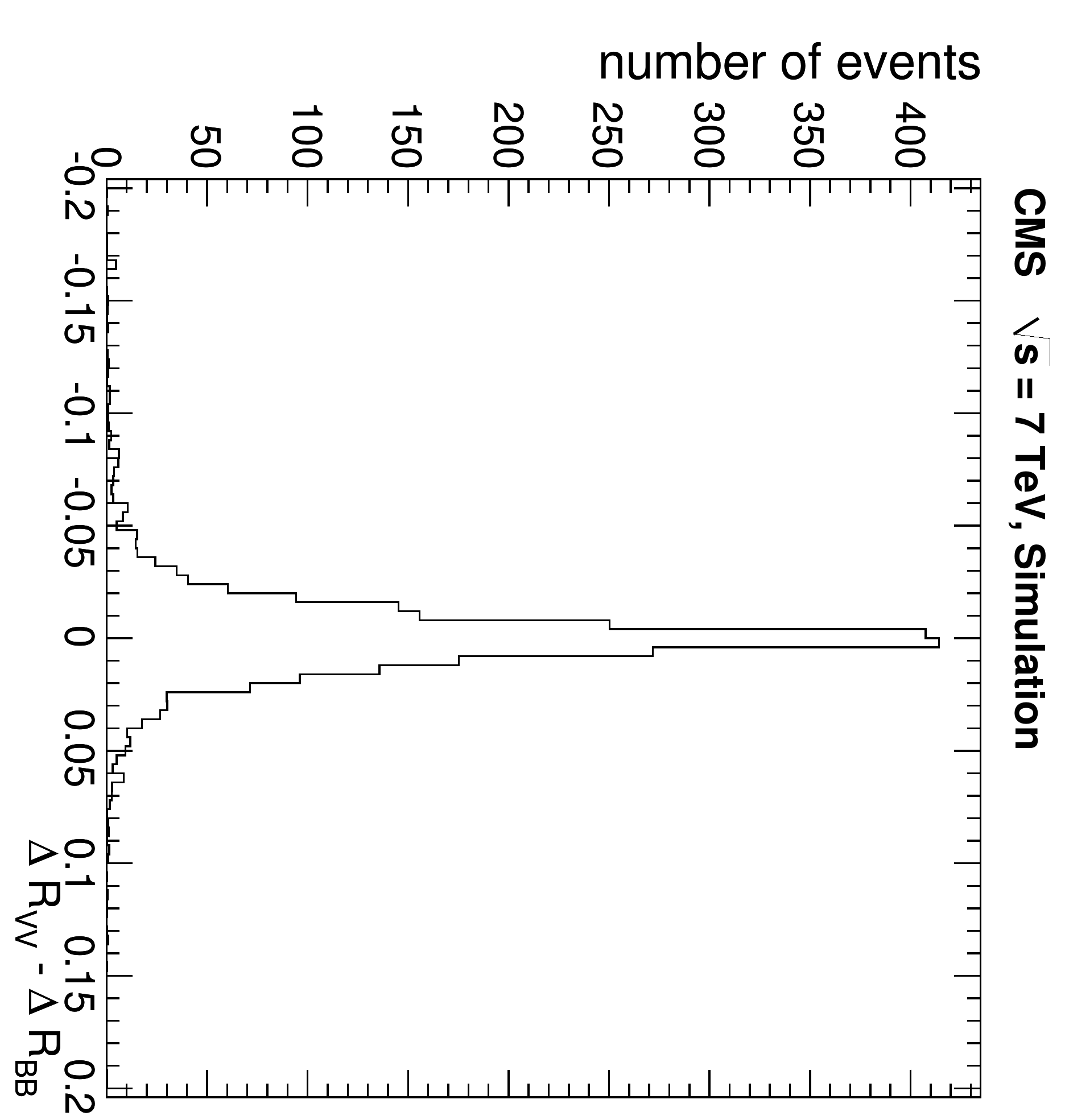}
\caption{ Resolution of the $\Delta R$ reconstruction, obtained using simulation
  for the leading jet $\pt > 84$ GeV sample.
  Left: $\Delta R$ values reconstructed between the two secondary vertices
   $\Delta R_{VV}$    versus the values between the original \PB\ hadrons $\Delta R_{BB}$,
   in the visible  \PB\ hadron phase space (see text). Right:
  projection onto the diagonal ($\Delta R_{VV} - \Delta R_{BB}$).
  The numbers in the boxes represent the number of events reconstructed in
  that particular bin.
}
\label{fig:deltar-reso}
\end{figure}

The resolution achieved in the $\Delta R$ reconstruction  is estimated
from simulation.
The comparison of the $\Delta R$ values reconstructed between
the two vertices $\Delta R_{VV}$ with the values calculated between the original
true \PB\ hadrons $\Delta R_{BB}$, determines the resolution.
This is illustrated in Fig.~\ref{fig:deltar-reso},
which shows the two-dimensional distribution  $\Delta R_{VV}$ versus $\Delta R_{BB}$
and its projection onto the diagonal ($\Delta R_{VV} - \Delta R_{BB}$).
A fit to this projection  directly yields an average
resolution better than 0.02 in $\Delta R$ for the core region, a value
much smaller than the  $\Delta R$ bin width of 0.4.

In order to calculate differential cross sections,
a $\Delta R$-dependent purity correction is applied.
The contributions to purity due to migration are illustrated in Fig.~\ref{fig:deltar-reso}a.
The total number of event entries off the diagonal is found to be about 3\%.
The largest impurity occurs close to $\Delta R_{VV} \approx 3$
as can be seen in the 2D plot. These events are due to misreconstructed
collinear events where only one \PB\  hadron is reconstructed,
while a fake vertex is found in the recoiling light quark jet.
The largest effect on a single bin is below 10\% and this is taken into account
in the purity correction.
The uncertainty arising from this correction is included in the systematic uncertainties.
The average \bhadb purity is found to be 84\%, with a variation within about $\pm 10\%$
over the full $\Delta R$ range
in the visible  region for the three leading jet \pt bins.

\subsection{Systematic Uncertainties}
\label{s:systematics}

Uncertainties relevant to the shape of the differential distributions are
crucial for this paper.
The consistency in shape between the data and the
simulation is assessed
and the systematic uncertainties are estimated by data driven methods.
The systematic uncertainties
related to the absolute normalisation are much larger than the shape dependent
ones.  They sum up to  a total of $47\%$,
 but do not affect the shape analysis (see below).
The dominant contribution originates from the \PB\  hadron reconstruction
efficiency ($\pm 20\%$, estimated in~\cite{ref:BTV-10-001}),
which amounts to a total of 44\% for reconstructing two \PB\  hadrons.

In the following the shape dependent systematic uncertainties
for the $\Delta R$ distributions are discussed.
The values are quoted in terms of the relative change of the
integrated cross section ratio $\rho_{\Delta R} = \sigma_{\Delta R < 0.8} / \sigma_{\Delta R >2.4}$.
Very similar systematic uncertainties arise for the
$\Delta \phi$ distributions and, hence, they are not quoted separately.

\begin{itemize}
\item {\it Algorithmic effects.}
The shape of the $\Delta R$ dependence of the efficiency $\alpha(\Delta R)$ is
checked by means of an event mixing method.
This event mixing technique mimics an event with two genuine SVs by merging
two independent events, where each has at least one reconstructed SV.
The positions of the two PVs are required to be within $20\, \mu{\rm m}$
in three-dimensional space.
This mixed event is then analysed and the fraction of cases where both
original SVs are again properly reconstructed is used to determine the
$\Delta{R}$ dependence of the efficiency to find two genuine SVs in an event
which had the SVs already reconstructed.
The shape of this efficiency $\alpha(\Delta R)$
is determined for the data and for the simulated samples independently in
bins of $\Delta{R}$.
The vertex reconstruction efficiency as a function of $\Delta R$
for data and for simulation, and their
ratio are shown in Fig.~\ref{fig:IVFratio}.
Since in this analysis the shape is the most relevant property,
the values in  Fig.~\ref{fig:IVFratio}b have been rescaled to the mean value.
This ratio exhibits good consistency in shape between simulation
and data over the full $\Delta R$ range, including the region of small $\Delta R$.
The differences are found to be within 2\%
and are taken as systematic uncertainties.

\begin{figure}[ht]
\centering
\includegraphics[angle=90,width=0.45\textwidth]{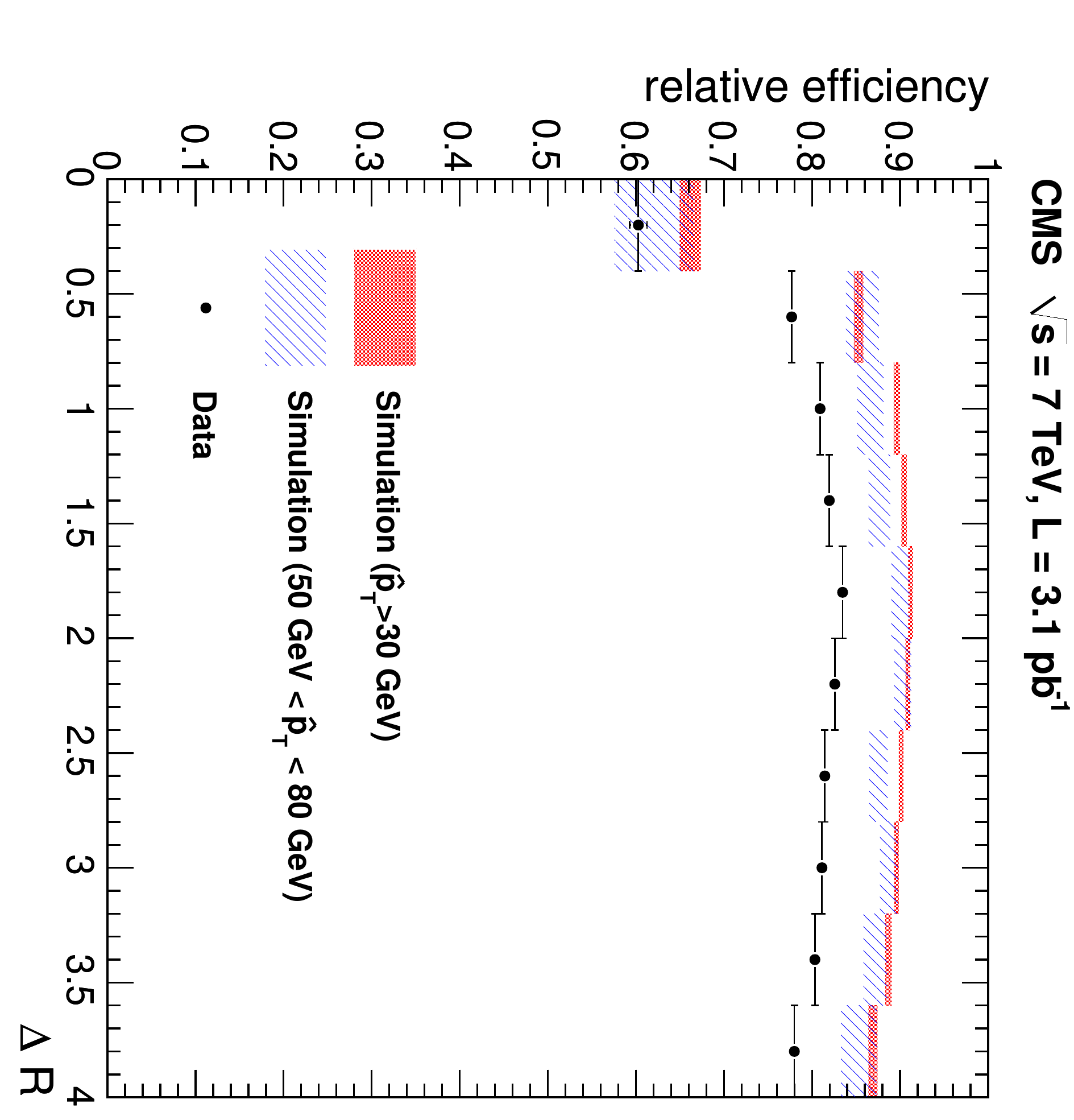}
\includegraphics[angle=90,width=0.45\textwidth]{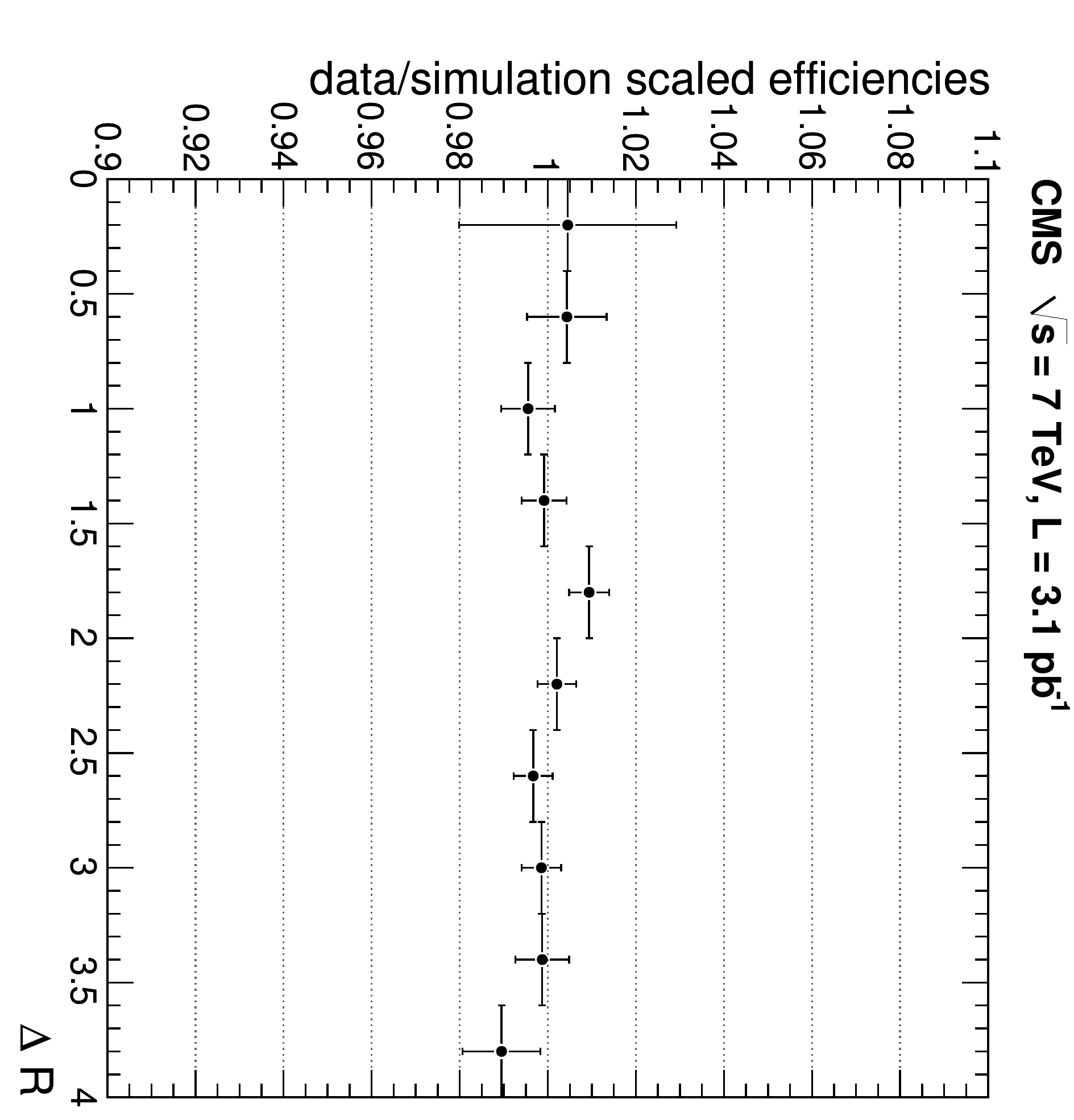}
\caption{Study of the vertex reconstruction efficiency by the event mixing method.
Shown as a function of $\Delta R$ are
the relative vertex reconstruction efficiency (left) $\alpha(\Delta R)$ (see text),
 and the ratio (right) between the quantities $\alpha(\Delta R)$
determined from the data and from the simulation.
The simulated $\alpha(\Delta R)$ distribution (left) is shown for two energy scales,
characterized by $\hat{p}_{\rm T}$,  the \PYTHIA parameter describing the transverse momentum
in the hard subprocess.
The ratio (right) is rescaled to unity to
estimate the accuracy of the simulated shape.
}
\label{fig:IVFratio}
\end{figure}

\item {\it \PB\  hadron momenta.}
The mean reconstruction efficiency for an observed $\Delta R$ value strongly
depends on the kinematic properties of the \PB\  hadron pair.
It depends on the \pt of each \PB\  hadron and predominantly
on the softer of the two.
Since all efficiency corrections are taken from the MC simulation,
it is important to verify that the kinematic behaviour
of the \bhadbbr pairs is also properly modelled by the simulation.
Confidence in the Monte Carlo modelling is provided by comparing the
transverse momentum distributions of the reconstructed \PB\  candidates
derived from data and from Monte Carlo simulation.
The distributions of the reconstructed \pt of the harder and of the softer
of the two hadrons, their asymmetry, as well as the
$\Delta R$ dependence of the average reconstructed \pt of the softer
hadron for the three leading jet \pt regions,
are shown in Fig.~\ref{fig:pt-asymmetry}.
The differences between the data and the simulation, convolved with the
\pt-dependent efficiency, are found to have an effect on the
final result of between 4\% and 8\%.
These values are used to estimate the systematic uncertainties reported
in Table~\ref{tab:relsyserr} as ``\PB\  hadron kinematics''.

\begin{figure}[htb]
	\centering
	\includegraphics[angle=90,width=0.45\textwidth]{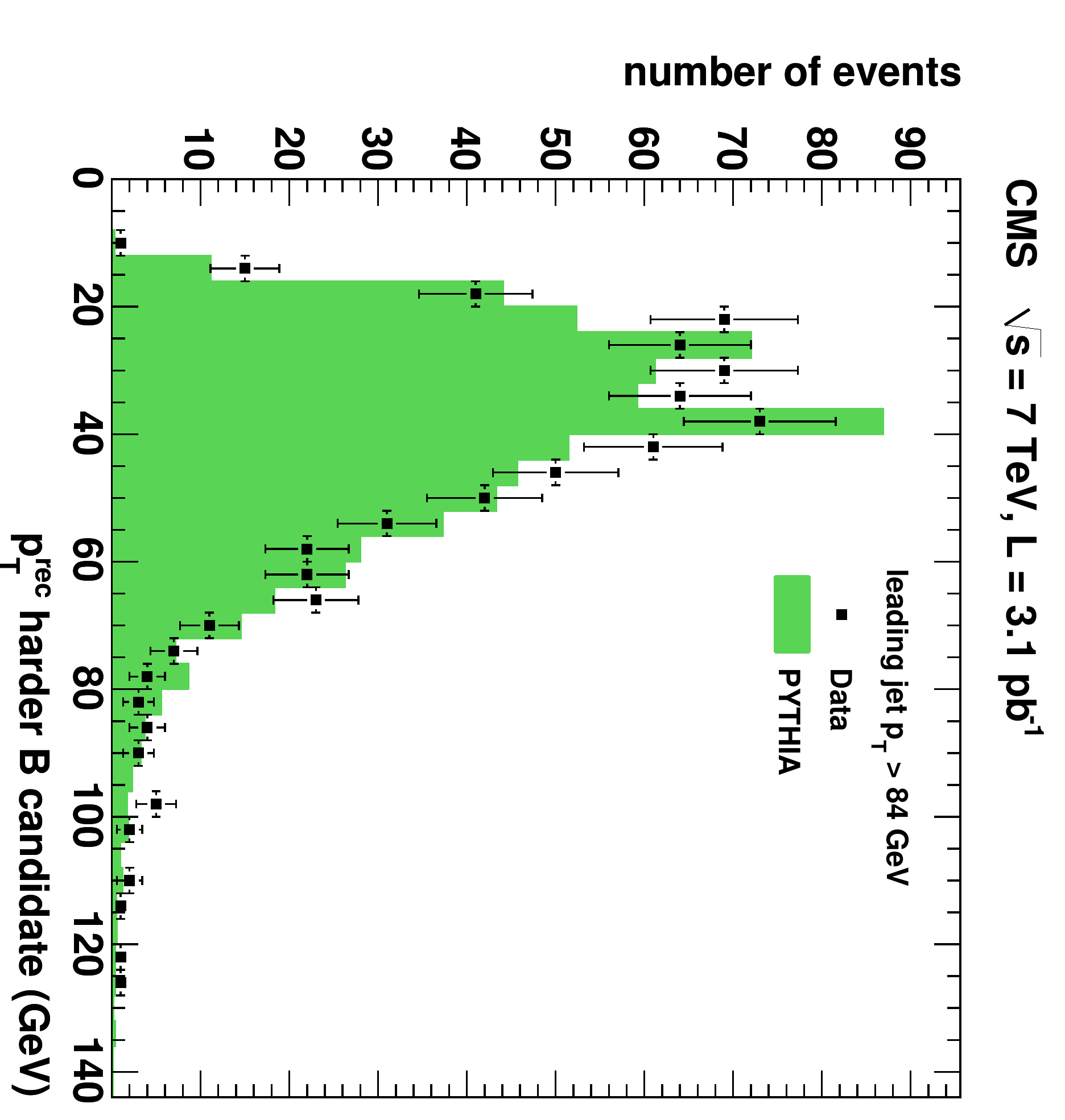}
	\includegraphics[angle=90,width=0.45\textwidth]{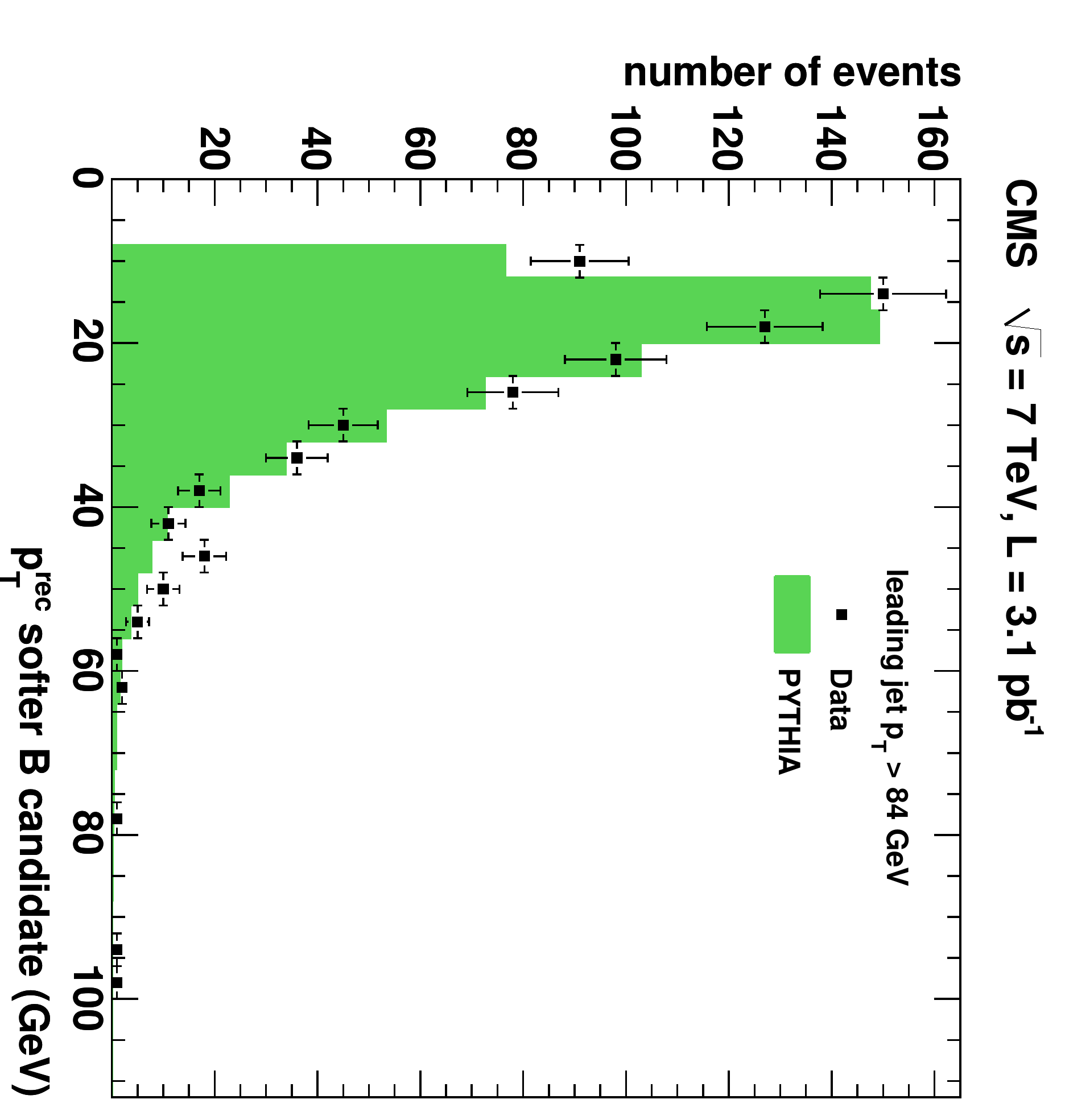} \\
	\includegraphics[angle=90,width=0.45\textwidth]{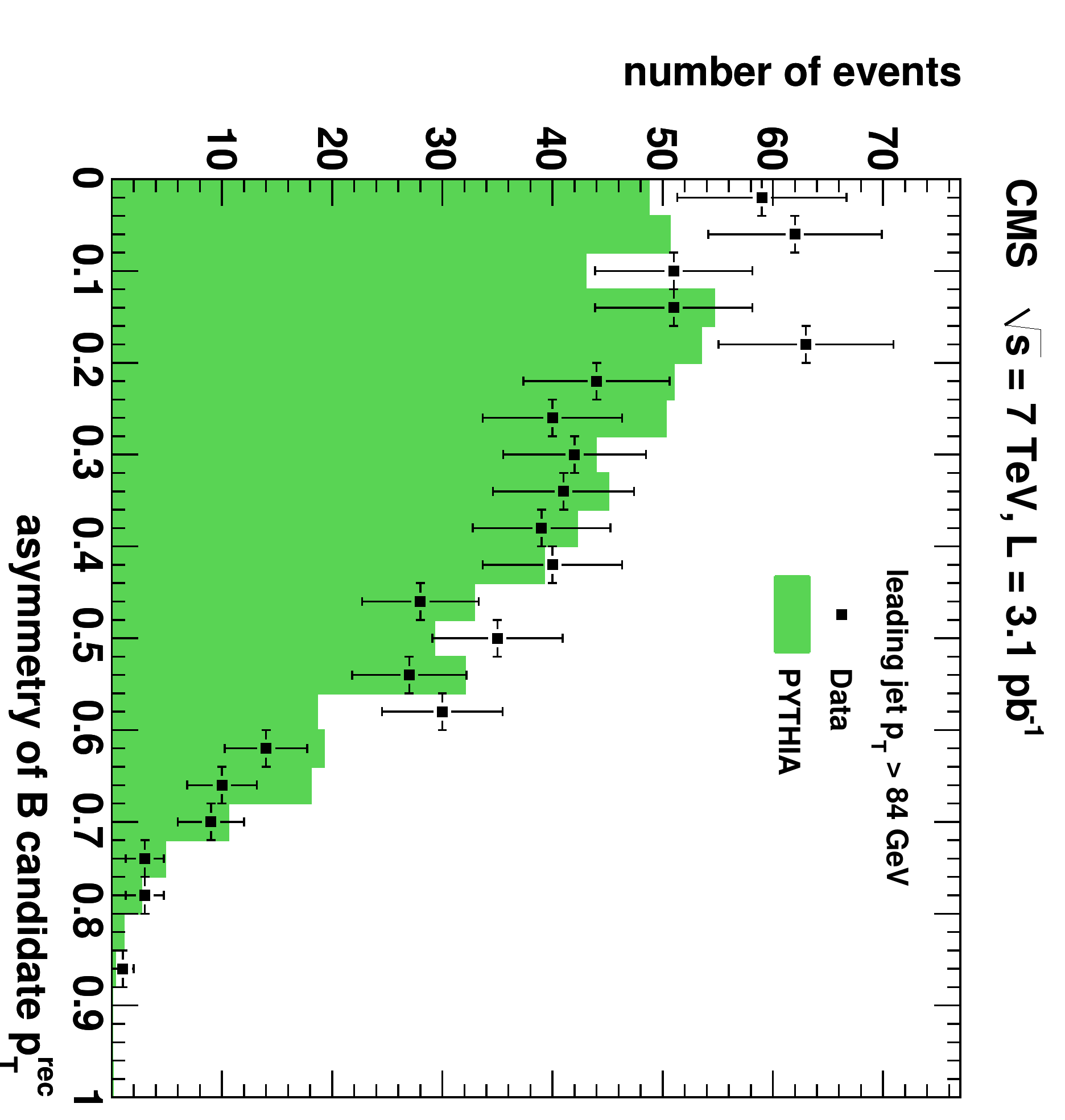}
	\includegraphics[angle=90,width=0.45\textwidth]{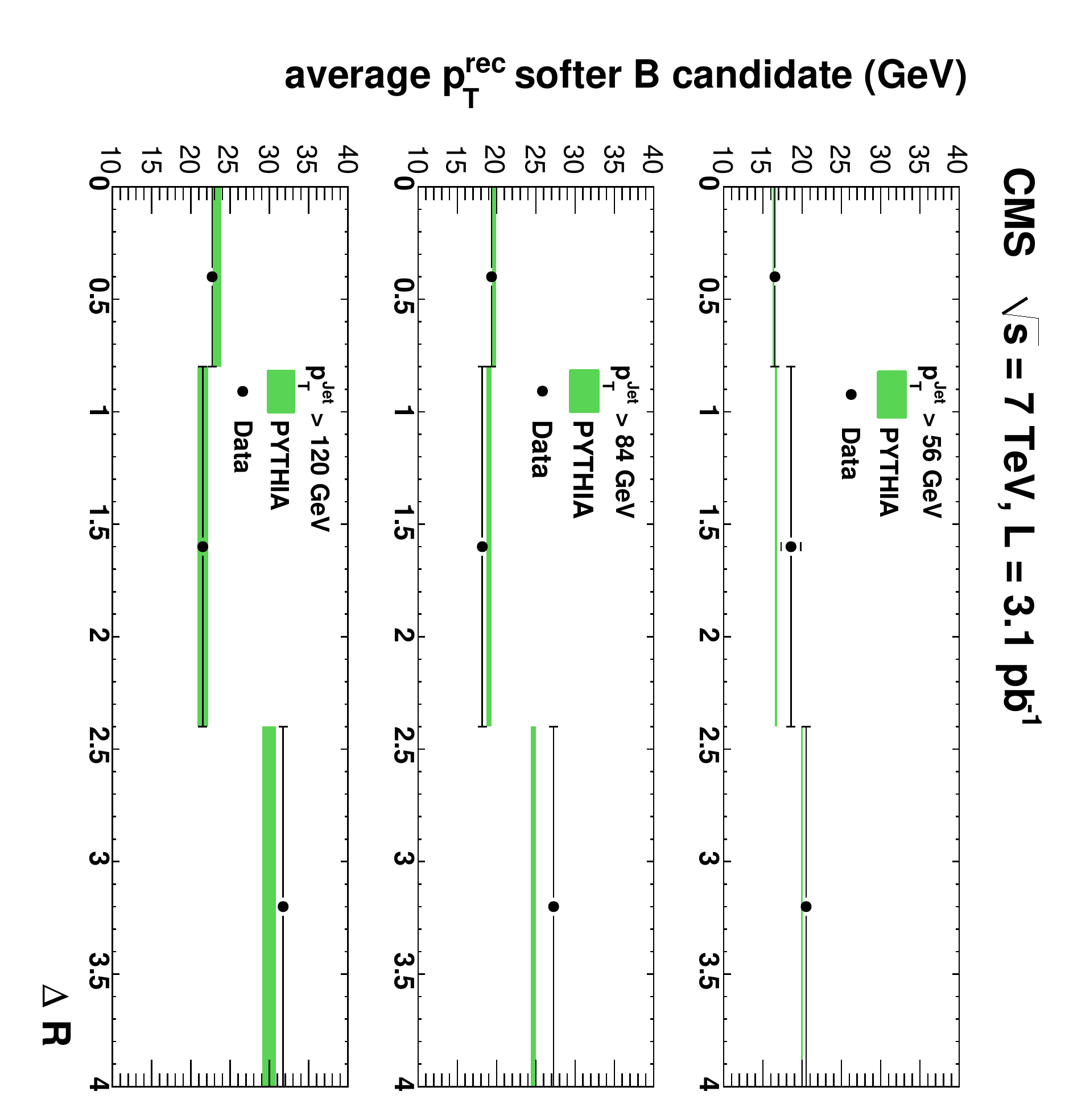}
	\caption{
          Distributions of the reconstructed \pt of the two \PB\  hadrons:
           \pt of the harder \PB\  hadron (top left);
          \pt of the softer \PB\  hadron (top right);
          asymmetry (bottom left) of the \pt of the harder and the softer \PB\  hadron;
           average \pt (bottom right) of the softer \PB\  hadron as a function of $\Delta R$
       for  data (solid dots) and {\PYTHIA} simulation (green bars) for the
       three leading jet \pt regions.
 }
\label{fig:pt-asymmetry}
\end{figure}

\item {\it Uncertainty on the Jet Energy Scale (JES).}
The JES influences the $\Delta R$ shape of the two \PB\ hadrons.
Its effect on the \pt of the leading jet is estimated assuming
a linear rise of the \pt dependency of the  relative cross section ratio
(see below).
Given that the higher \pt scales exhibit a larger
relative contribution to the cross section at low $\Delta R$,
the actual $\Delta R$ shape is distorted by this effect.
The uncertainty on the JES is determined by assuming a
$\pm 3\%$~\cite{CMS-PAS-JME-10-010} uncertainty on average for the energy region relevant for this analysis.
An additional $\pm 5\%$ is added to take into account the differences in the
jet energy corrections between \Pbquark and light jets as estimated in the simulation.
This yields a variation in the $\Delta R$ shape within 6\%,
which is  taken as systematic uncertainty.

\item {\it Phase space correction.}
The measurements of vertices are corrected to the visible
phase space of the \PB\ hadrons defined by $|\eta(\mathrm{B})| < 2.0$ and $\pt(\mathrm{B}) > 15$ GeV,
using the {\PYTHIA} Monte Carlo simulation.
In the analysis only reconstructed \PB\ hadrons
above a \pt of  8 GeV are considered.
The uncertainty arising from this choice has been estimated
by varying the \pt cut on the reconstructed vertex from 8 to 10 GeV,
recomputing the MC correction and repeating the final measurement.
The uncertainty is  found to be 2.8\%.

\item {\it Migration.} The bin-to-bin migrations  in the  sample are small because,
 as shown in Fig.~\ref{fig:deltar-reso},
 the core of the vertex resolution in $\Delta R$ (0.02) is
  much smaller than the chosen bin width (0.4).
  The migrations are taken into account through the efficiency corrections.
  The off-diagonal contributions (predominantly at $\Delta R_{VV} \approx \pi$ from
  misreconstructed collinear gluon splitting events,  with one vertex from the
  recoiling jet)
  are subtracted on a bin-to-bin basis.
   An uncertainty of up to 2.1\% on this purity correction is obtained by increasing
   the small angle $\Delta R<0.8 $ contribution by 50\%
   (compatible with the measured results, as presented below).

\item {\it Monte Carlo statistics.}
   An additional bin-to-bin systematic uncertainty results from the limited
   number of simulated events.
   An uncertainty of 13\% is used, conservatively taking
   the maximum value of either the statistical uncertainty
   of the simulation or half of the largest bin-to-bin fluctuation observed in
   the correction function between any of the $\Delta R$ bins.
   This uncertainty is mostly relevant for Figs. \ref{fig:b2b}
   and \ref{fig:ratio_b2b}; its effect is reduced in
   Fig.~\ref{fig:gsp_fcr}.
\end{itemize}

\begin{table}[htb]
\caption{ Systematic uncertainties affecting the shape of the
  differential cross section as a function of $\Delta R$, for the
  three leading jet \pt regions.
  The values are quoted in terms of percentage changes
  of  the integrated cross section ratio
 $\rho_{\Delta R}$.
  In the figures, these values are included for each bin.
  Very similar systematic uncertainties are assumed for the
  $\Delta \phi$ distributions.}

 \renewcommand{\arraystretch}{1.4}
 \centering
 \begin{tabular}{|c|c|c|c|}
  \hline
  {Source of uncertainty in shape} &
  \multicolumn{3}{c|}{Change in $\rho_{\Delta R} = \sigma_{\Delta R < 0.8} / \sigma_{\Delta R >2.4}$ (\%)} \\
\hline
 & \multicolumn{3}{c|}{Leading jet $\pt$ bin (GeV)}\\
 & $> 56$  & $> 84$ & $> 120$ \\
\hline
\hline
 Algorithmic effects (data mixing)  & $ 2.0$   & $ 2.0$   & $ 2.0$    \\
 \PB\  hadron kinematics (\pt of softer \PB)  & $ 8.0$    & $ 7.0$    & $ 4.0$    \\
 Jet energy scale       & $ 6.0$   & $ 6.0$ & $ 6.0$   \\
 Phase space correction & $ 2.8$   & $ 2.8$  & $ 2.8$ \\
 Bin migration from resolution    & $ 0.6$  & $ 1.3$ & $ 2.1$  \\
\hline
\hline
Subtotal shape uncertainty  & $ 10.6$   & $ 9.9$   & $ 8.3$   \\
\hline
 MC statistical uncertainty  & $ 13.0$  & $ 13.0$ & $ 13.0$  \\
\hline
\hline
Total shape uncertainty & $ 16.8$  & $ 16.4$ & $ 15.4$  \\
\hline
 \end{tabular}
\label{tab:relsyserr}
\end{table}

The shape-dependent systematic uncertainties are calculated
and included binwise in the figures, as indicated by the outer error bars
which show statistical and systematic uncertainties added in quadrature.
They are summarised in Table~\ref{tab:relsyserr}.
The overall normalisation uncertainties are not included
in the error bars in the figures.

\section{Results}
\label{s:results}

\subsection{Differential Distributions in $\Delta R$ and $\Delta \phi$}

The differential cross section of \bhadbbr-pair production
is measured as a function of the  angular separation variables
$\Delta R$ and $\Delta \phi$ between the two reconstructed \PB\  hadrons
for three different energy scales.
The results are presented for the visible kinematic phase space
of the  \PB\  hadrons
and the leading jet \pt ranges as defined in Section~\ref{s:analysis}.
The cross sections are determined by applying efficiency corrections
and normalising to the total
integrated luminosity, according to

\begin{equation}
\left( \frac{d\sigma_{\rm visible}(p p \to \bhadb \ X)}{dA}\right)_i =
 \frac{N_i(\mathrm{data}) \cdot f_{i}}{\Delta A_i \cdot {\cal L} \cdot \epsilon_i} \, ,
\label{eq:exp-sigma}
\end{equation}

where $N_i(\mathrm{data})$ denotes the number of selected signal \bhadb events
in bin $i$, ${\cal L}$  the integrated luminosity,
$\epsilon_i$ the total efficiency, $f_{i}$ the purity correction factor,
and $\Delta A_i$ the width of bin $i$ in variable $A$,
with $A$ being  $\Delta R$ or $\Delta \phi$.

\begin{figure}[htb]
	\centering
\includegraphics[width=0.45\textwidth]{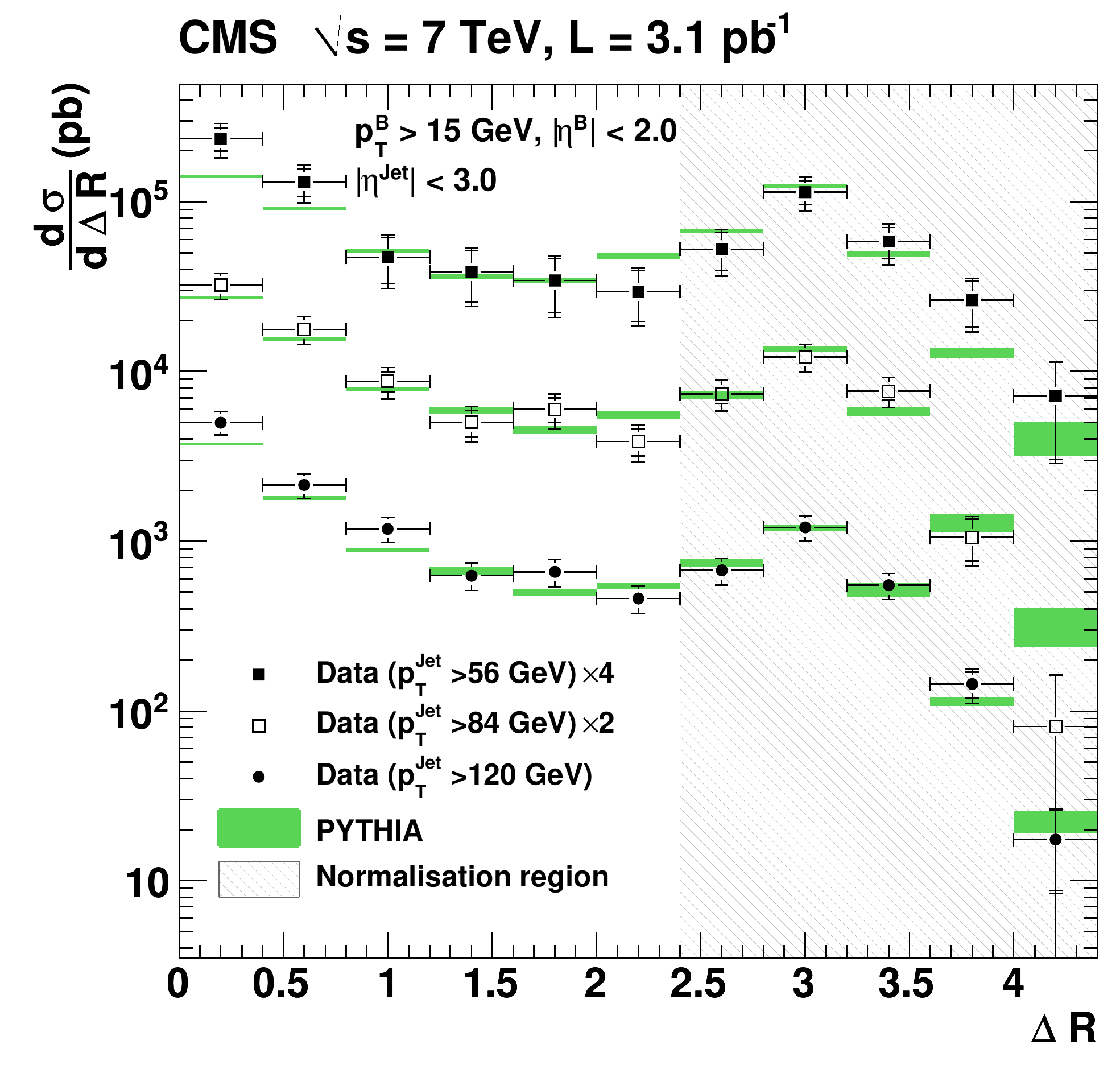}
\includegraphics[width=0.45\textwidth]{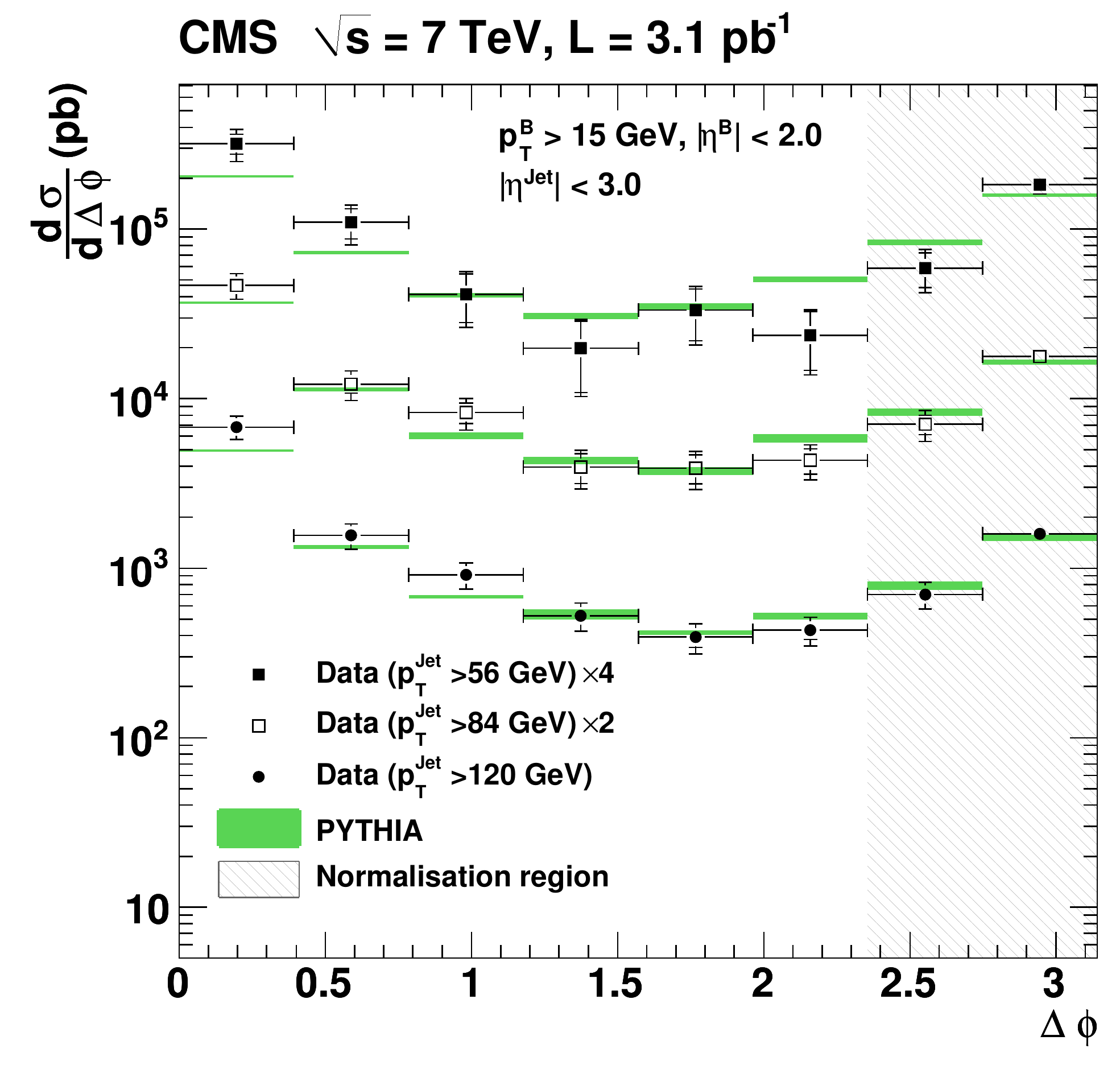}
	\caption{Differential \bhadb production cross sections as a function of
         $\Delta R$ (left) and $\Delta \phi$ (right) for the three leading jet \pt regions.
   For clarity, the $\pt > 56$ and 84 GeV bins are offset by a factor 4 and 2,
   respectively.
   For the data points, the error bars show the statistical (inner bars) and the
   total (outer bars) uncertainties.
   A common uncertainty  of 47\% due to the absolute normalisation on
   the data points is not included.
   The symbols denote the values averaged over the bins and
   are plotted at the bin centres.
   The {\PYTHIA} simulation (shaded bars) is normalised to the region $\Delta R > 2.4$
   or $\Delta \phi > 2.4$,  as indicated
   by the shaded normalisation regions.
   The widths of the shaded bands indicate the statistical uncertainties of the predictions.
}
\label{fig:b2b}
\end{figure}

The measured cross sections are shown in Fig.~\ref{fig:b2b}
as a function of $\Delta R$ and $\Delta \phi$ for the three
leading jet \pt regions. The error bars on the data points include statistical
and uncorrelated systematic uncertainties.
An uncertainty of 47\% common to all data points
due to the absolute normalisation is not shown in the figure.
The bars shown for the \PYTHIA simulation
in Fig.~\ref{fig:b2b} are normalised to the
region $\Delta R > 2.4$ or $\Delta \phi > 2.4$,
where the theory calculations are expected to be more reliable,
since the cross section is anticipated to be dominated by
leading order diagrams (flavour creation).

\begin{figure}[htb]
\centering
\includegraphics[angle=90,width=0.45\textwidth]{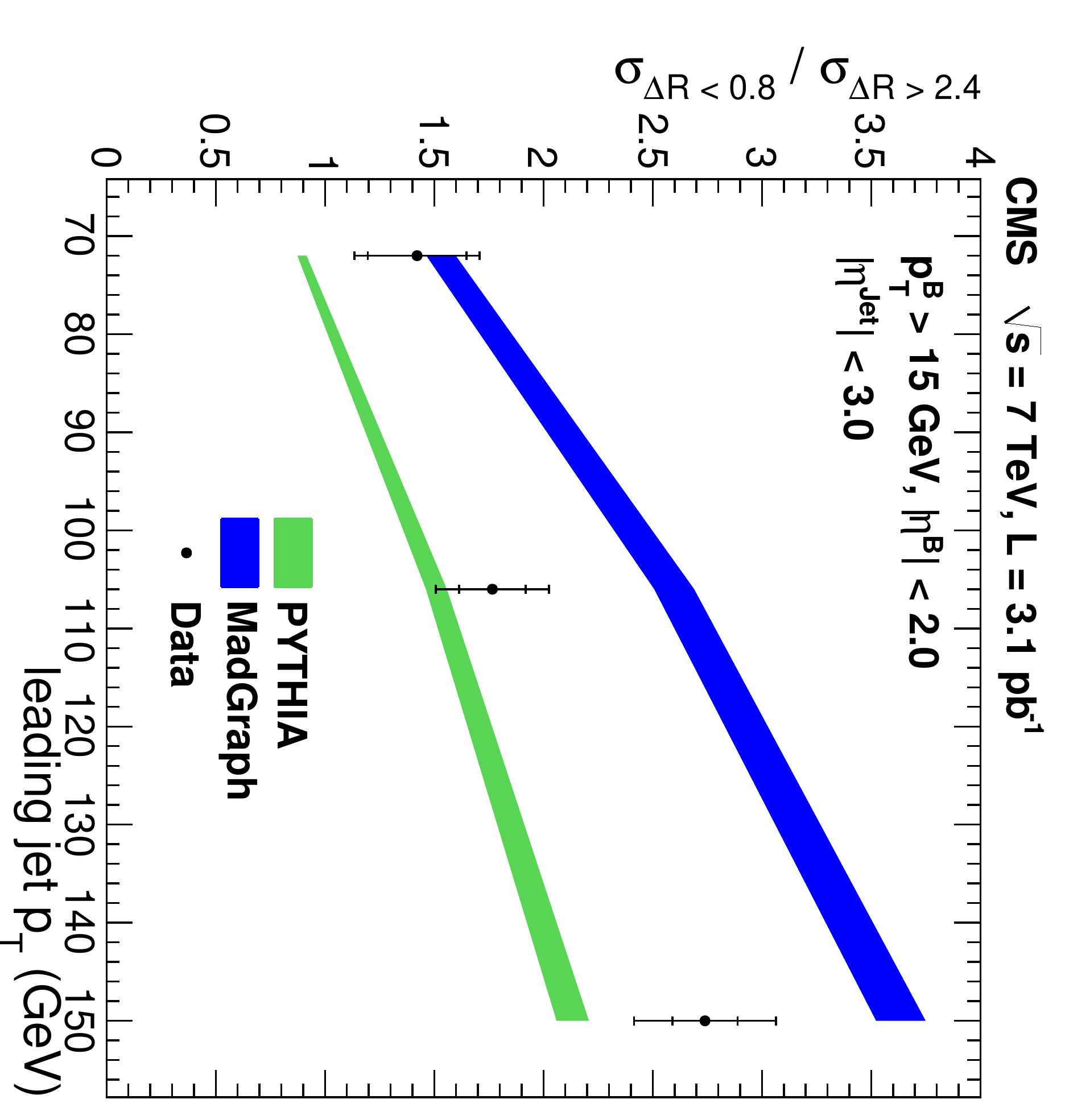}
\includegraphics[angle=90,width=0.45\textwidth]{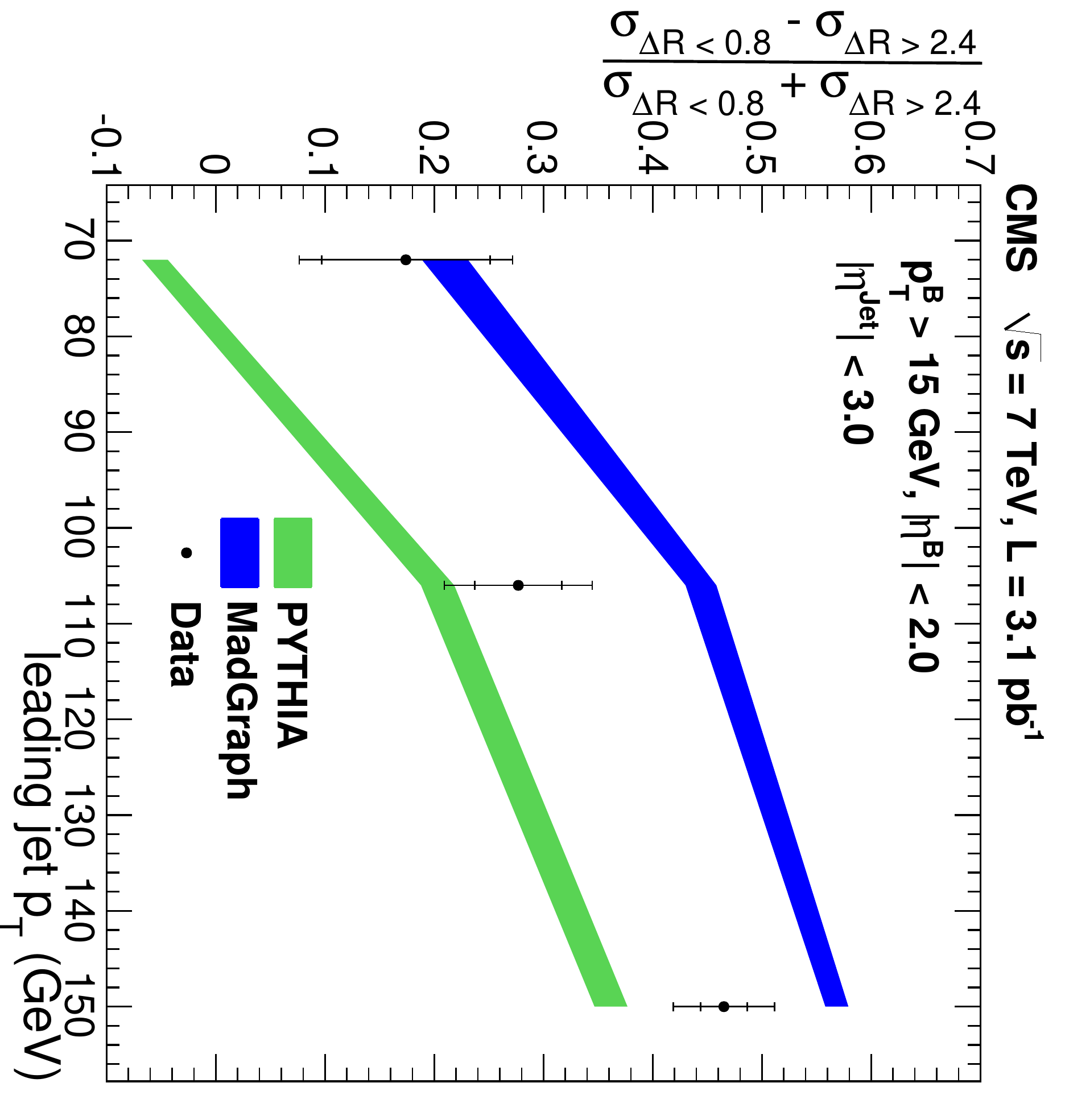}
\caption{Left: ratio between the \bhadb production cross sections in
   $\Delta R < 0.8$ and $\Delta R>2.4$, $\rho_{\Delta R}=\sigma_{\Delta R < 0.8}$ / $\sigma_{\Delta R >2.4}$,
   as a function of the leading jet \pt. Right: asymmetry between the two regions,
$(\sigma_{\Delta R < 0.8} - \sigma_{\Delta R >2.4})$ /
$(\sigma_{\Delta R < 0.8} + \sigma_{\Delta R >2.4})$.
   The symbols denote the data averaged over the bins and
   are plotted at the mean leading jet \pt of the bins.
   For the data points, the error bars show the statistical (inner bars) and the
   total (outer bars) errors.
   Also shown are the predictions from the {\PYTHIA} and {\MADGRAPH} simulations,
   where the widths of the bands indicate the uncertainties arising from the
   limited number of simulated events.
    }
\label{fig:gsp_fcr}
\end{figure}

It is interesting to note that the cross sections at small
values of $\Delta R$ or $\Delta \phi$ are found to be substantial.
They exceed the cross sections observed at large angular separation values,
the configuration where the two \PB\ hadrons are
emitted in opposite directions.

The scale dependence is illustrated in Table~\ref{tab:gsp-fcr-ratio} and Fig.~\ref{fig:gsp_fcr},
where the left panel shows the ratio $\rho_{\Delta R}$ as a function of the leading jet \pt,
a measure of the hard interaction scale.
The right panel shows the asymmetry of the cross section
contributions between small and large $\Delta R$  values,
$(\sigma_{\Delta R < 0.8} - \sigma_{\Delta R >2.4})$ /
$(\sigma_{\Delta R < 0.8} + \sigma_{\Delta R >2.4})$.
The measured data clearly indicate that the
relative contributions of $\sigma_{\Delta R < 0.8}$ significantly exceed those of
$\sigma_{\Delta R >2.4}$.
In addition, the data show that this excess
depends on the energy scale,
increasing towards larger leading jet \pt values.

\begin{table}[htb]
\caption{\pt cut of the leading jet, average jet \pt, cross sections
  in the two $\Delta R$ regions (including the $47\%$ uncertainty on the
  absolute normalisation), average efficiency, average purity,
  and cross section ratio for the data, as well as for the {\PYTHIA} and {\MADGRAPH} simulations.
  Statistical and systematic uncertainties are included for the data, while for
  the simulations only the statistical uncertainties are given.
}

 \centering
 \begin{tabular}{|c|c|c|c|c|c|c|c|c|}
  \hline
  \multicolumn{2}{|c}{Jet \pt} &
  \multicolumn{4}{|c}{} &
  \multicolumn{3}{|c|}{$ \rho_{\Delta R} = \sigma_{\Delta R < 0.8}\, /\, \sigma_{\Delta R >2.4} $} \\
\hline
  Cut & $\langle\pt\rangle$ &$\sigma_{\Delta R < 0.8}$ & $\sigma_{\Delta R >2.4}$ &
 $\langle\epsilon\rangle$ & $\langle P \rangle$ & Data & {\PYTHIA}  & {\MADGRAPH}   \\
  (GeV) & (GeV) & (nb) & (nb) & ($\%$) & ($\%$) & (stat+sys) & (stat) & (stat)   \\
\hline
  $> 56$  & 72  & $37 \pm 26 $ & $26\pm 16 $ & $7.4$  & $84.9$ & $ 1.42\pm 0.29$ &
  $0.89\pm 0.02$ & $1.53\pm 0.07$ \\
  $> 84$  & 106 & $10 \pm 4 $ & $5.6\pm 4.0 $ & $9.3$  & $84.6$ & $ 1.77\pm 0.26$ &
  $1.51\pm 0.05$ & $2.60\pm 0.09$ \\
  $> 120$  & 150  & $2.8 \pm 1.0 $ & $1.0\pm 1.2 $ & $10.7$  & $83.2$ & $ 2.74\pm 0.32$ &
  $2.13\pm 0.07$ & $3.64\pm 0.11$ \\
\hline
 \end{tabular}
\label{tab:gsp-fcr-ratio}
\end{table}

\subsection{Comparisons with Theoretical Predictions}

The measured distributions are compared with various theoretical
predictions, based on perturbative QCD calculations, both at LO and NLO.

Within pQCD, a back-to-back configuration
for the production of the \bhadbbr pair
(i.e.\ large values of  $\Delta R$ and/or  $\Delta \phi$) is expected
for the LO processes, while
the region of phase space with small opening angles between
the \PB\  and $\overline{\rm B}$ hadrons provides strong sensitivity to
collinear emission processes.
The higher-order
processes, such as gluon radiation which splits into
\Pbbbar pairs, are anticipated to have a smaller angular
separation between the \Pbquark quarks.
Naively, the flavour creation contribution is expected to be
dominant in most regions of the phase space, whereas the
gluon splitting contributions should be relatively small.

\begin{figure}[htb]
	\centering
\includegraphics[width=0.45\textwidth]{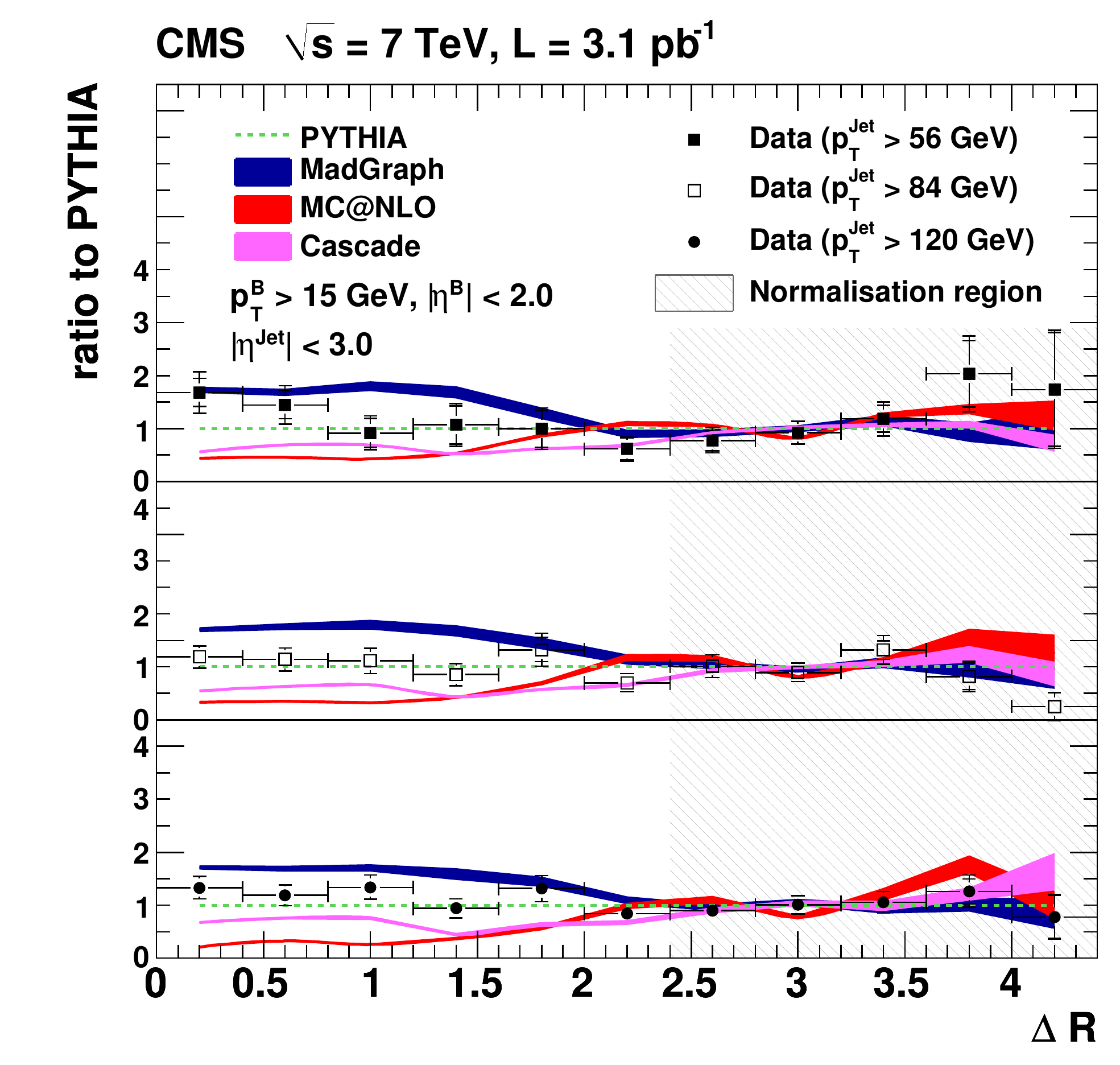}
\includegraphics[width=0.45\textwidth]{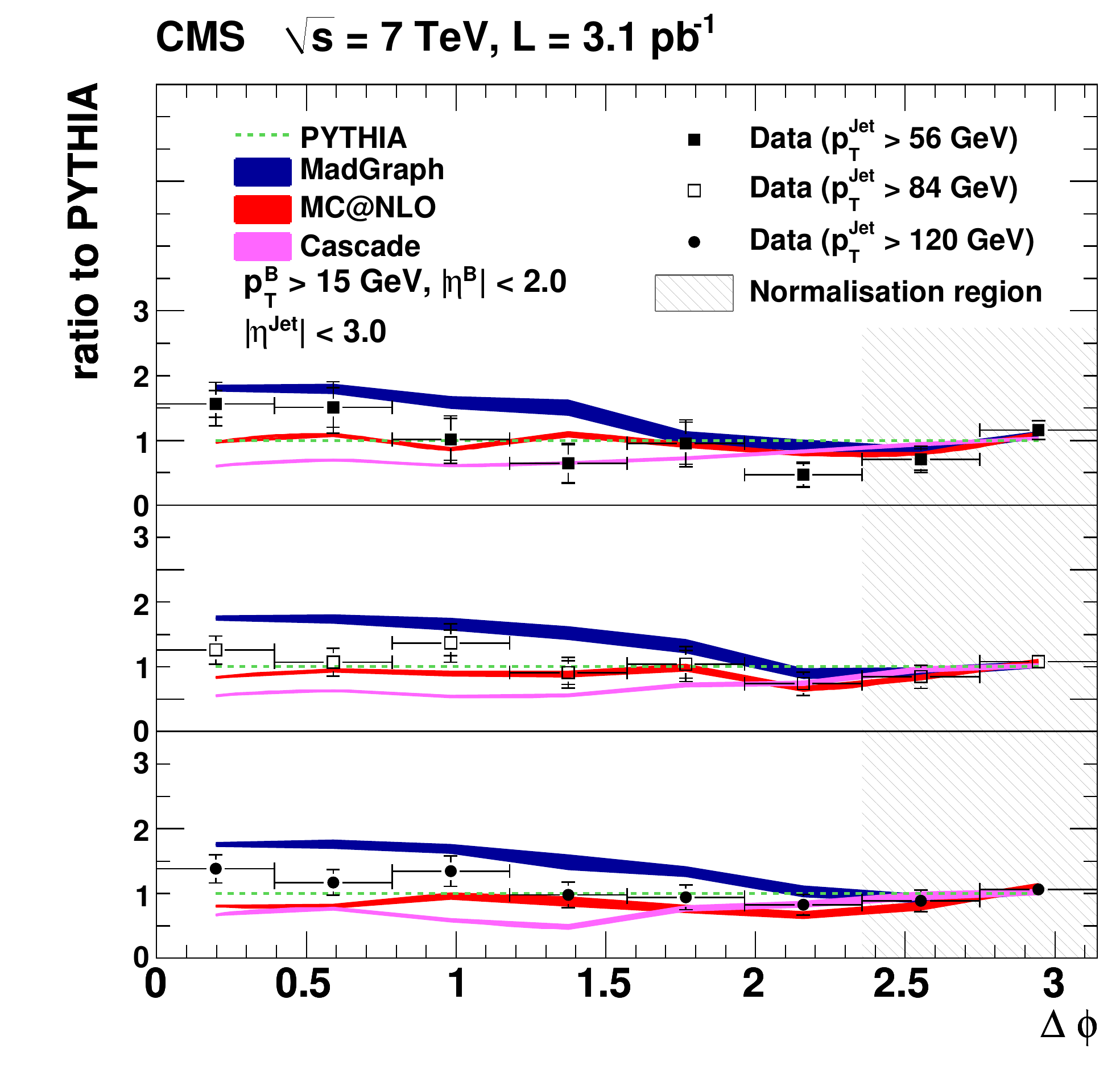}
	\caption{Ratio of the differential \bhadb production cross sections, as a function of
         $\Delta R$ (left) and $\Delta \phi$ (right),
         for data, \MADGRAPH, \MCATNLO and \CASCADE, with respect to
         the {\PYTHIA} predictions, for the three leading jet \pt bins.
        The simulation is normalised to the region $\Delta R > 2.4$
        and $\Delta \phi > 2.4$ (FCR region), as indicated
        by the shaded normalisation region.
        The widths of the theory bands
        indicate the statistical uncertainties of the simulation.
}
\label{fig:ratio_b2b}
\end{figure}

The measurements show that the \bhadb production cross section ratio
$\rho_{\Delta R}$ increases as a function of the
leading jet \pt in the event (see Fig.~\ref{fig:gsp_fcr}).
Larger \pt values lead to more gluon radiation and, hence, are
expected to produce more gluon splitting into \bhadbbr pairs.
This general trend is described by the theoretical calculations.

In order to provide a detailed comparison between the data and the theory
predictions in terms of shape, Fig.~\ref{fig:ratio_b2b} presents the ratios, of
the data as well as of the {\MADGRAPH}, {\MCATNLO} and {\CASCADE}
models, with respect to the {\PYTHIA} predictions,
for the three different scales in leading jet \pt.
The values for the {\PYTHIA} simulation are normalised
in the region $\Delta R>2.4$ (or $\Delta \phi>2.4$).

It is observed that none of the predictions describes the data very well.
The data lie between the {\MADGRAPH} and the {\PYTHIA} curves.
The \MCATNLO calculations do not describe the shape of the
observed $\Delta R$ distribution.
In particular, at small values of $\Delta R$,
where higher-order processes, notably gluon splitting,
are expected to be large, the {\MCATNLO} predictions are substantially
below the data.
The $\Delta \phi$ distribution is more adequately reproduced by {\MCATNLO}.
The {\CASCADE} predictions are significantly below the data in all regions,
both in the $\Delta R$ and $\Delta \phi$ distributions.

\section{Summary}
\label{s:conclusions}

A first measurement of the angular correlations between \bhadbbr pairs produced
in pp collisions at a centre-of-mass energy of $ 7\; \mbox{TeV}$  is presented.
The measurements are based on data corresponding to an integrated luminosity of
$3.1\, \pm 0.3\,  \mathrm{pb}^{-1}$ recorded by the CMS experiment during 2010.
The detection of the \PB\  hadrons is based on the reconstruction of the secondary
vertices from their decays.
The results are given in terms of normalised differential production cross sections
as functions of the angular separation variables $\Delta R$ and $\Delta \phi$
between the two \PB\  hadrons.
The data exhibit a substantial enhancement of the cross section at
small angular separation, exceeding the values measured at large
$\Delta R$ and $\Delta \phi$.
The fraction of cross section in this collinear region is found to increase
with the leading jet \pt of the event.

The measurements are compared to predictions, based on LO and NLO perturbative QCD
calculations.
Overall, it is found that the data lie between the {\MADGRAPH}
and the {\PYTHIA} predictions.
Neither the {\MCATNLO} nor the {\CASCADE} calculations
describe the shape of the $\Delta R$ distribution well.
In particular the collinear region at small values of $\Delta R$, where the
contributions of gluon splitting processes are expected to be large,
is not adequately described by any of the predictions.  \\

\section*{Acknowledgements}
We wish to congratulate our colleagues in the CERN accelerator departments for the excellent
performance of the LHC machine.
We thank the technical and administrative staff at CERN and
other CMS institutes, and acknowledge support from:
FMSR (Austria); FNRS and FWO (Belgium);
CNPq, CAPES, FAPERJ, and FAPESP (Brazil); MES (Bulgaria); CERN; CAS, MoST,
and NSFC (China); COLCIENCIAS (Colombia); MSES (Croatia); RPF (Cyprus); Academy of Sciences
and NICPB (Estonia); Academy of Finland,ME, andHIP (Finland); CEAand CNRS/IN2P3 (France);
BMBF, DFG, and HGF (Germany); GSRT (Greece); OTKA and NKTH (Hungary);
DAE and DST (India); IPM (Iran); SFI (Ireland); INFN (Italy); NRF and WCU (Korea); LAS
(Lithuania); CINVESTAV, CONACYT, SEP, and UASLP-FAI (Mexico); PAEC (Pakistan); SCSR
(Poland); FCT (Portugal); JINR (Armenia, Belarus, Georgia, Ukraine, Uzbekistan); MST and
MAE (Russia); MSTD (Serbia); MICINN and CPAN (Spain); Swiss Funding Agencies
(Switzerland); NSC (Taipei); TUBITAK and TAEK (Turkey); STFC (United Kingdom);
DOE and NSF (USA).

\clearpage

\bibliography{auto_generated}   

\providecommand{\href}[2]{#2}\begingroup\raggedright\begin{thebibliography}{10}%
\makeatletter
\providecommand{\hrefCMSnoop }[0]{\@secondoftwo}%
\makeatother

\bibitem{CMS:2011hf}
\hrefCMSnoop {} {{ CMS} Collaboration, ``{Inclusive b-hadron Production Cross
  Section with Muons in pp Collisions at sqrt(s) = 7 TeV}'',}
\href{http://www.arXiv.org/abs/1101.3512}{\texttt{ arXiv:1101.3512}}.

\bibitem{ref:BPH-10-009}
\hrefCMSnoop {} {{ CMS} Collaboration, ``{Inclusive $b$-Jet Production in pp
  Collisions at $\sqrt{s}=7\,$ TeV}'',} \textit{ CMS Note} \textbf{
  CERN-CMS-PAS-BPH-10-009} (2010).

\bibitem{Aaij:2010gn}
\hrefCMSnoop {} {{ LHCb} Collaboration, ``{Measurement of $\sigma(pp
  \rightarrow b\bar{b} X $) at $\sqrt{s}=7\TeV$ in the forward region}'',}
  \textit{ Phys. Lett.} \textbf{ B694} (2010) 209,
  \href{http://www.arXiv.org/abs/1009.2731}{\texttt{ arXiv:1009.2731}}.
\href{http://dx.doi.org/10.1016/j.physletb.2010.10.010}{\texttt{
  doi:10.1016/j.physletb.2010.10.010}}.

\bibitem{ref:BTV-10-001}
\hrefCMSnoop {} {{ CMS} Collaboration, ``{Commissioning of $b$ Jet
  Identification with pp Collisions at $\sqrt{s}=7\,$ TeV}'',} \textit{ CMS
  Note} \textbf{ CERN-CMS-PAS-BTV-10-001} (2010).

\bibitem{Acosta:2004nj}
\hrefCMSnoop {} {{ CDF} Collaboration, ``{Measurements of $b\bar{b}$ azimuthal
  production correlations in $p\bar{p}$ collisions at $\sqrt{s} = 1.8$ TeV}'',}
  \textit{ Phys. Rev.} \textbf{ D71} (2005) 092001,
  \href{http://www.arXiv.org/abs/hep-ex/0412006}{\texttt{
  arXiv:hep-ex/0412006}}.
\href{http://dx.doi.org/10.1103/PhysRevD.71.092001}{\texttt{
  doi:10.1103/PhysRevD.71.092001}}.

\bibitem{Aaltonen:2007zza}
\hrefCMSnoop {} {{ CDF} Collaboration, ``{Measurement of correlated $b\bar{b}$
  production in $p\bar{p}$ collisions at $\sqrt{s}$ = 1960 GeV}'',} \textit{
  Phys. Rev.} \textbf{ D77} (2008) 072004,
  \href{http://www.arXiv.org/abs/0710.1895}{\texttt{ arXiv:0710.1895}}.
\href{http://dx.doi.org/10.1103/PhysRevD.77.072004}{\texttt{
  doi:10.1103/PhysRevD.77.072004}}.

\bibitem{cms:2008zzk}
\hrefCMSnoop {} {{ CMS} Collaboration, ``{The CMS experiment at the CERN
  LHC}'',} \textit{ JINST} \textbf{ 3} (2008) S08004.
\href{http://dx.doi.org/10.1088/1748-0221/3/08/S08004}{\texttt{
  doi:10.1088/1748-0221/3/08/S08004}}.

\bibitem{Zanderighi:2007dy}
\hrefCMSnoop {} {G.~Zanderighi, ``{Accurate predictions for heavy quark
  jets}'',} \href{http://www.arXiv.org/abs/0705.1937}{\texttt{
  arXiv:0705.1937}}.
\href{http://dx.doi.org/10.3360/dis.2007.180}{\texttt{
  doi:10.3360/dis.2007.180}}.

\bibitem{Banfi:2007gu}
\hrefCMSnoop {} {A.~Banfi, G.~P. Salam, and G.~Zanderighi, ``{Accurate QCD
  predictions for heavy-quark jets at the Tevatron and LHC}'',} \textit{ JHEP}
  \textbf{ 07} (2007) 026, \href{http://www.arXiv.org/abs/0704.2999}{\texttt{
  arXiv:0704.2999}}.
\href{http://dx.doi.org/10.1088/1126-6708/2007/07/026}{\texttt{
  doi:10.1088/1126-6708/2007/07/026}}.

\bibitem{Sjostrand:2006za}
\hrefCMSnoop {} {T.~Sj{\"o}strand, S.~Mrenna, and P.~Z. Skands, ``{PYTHIA 6.4
  Physics and Manual}'',} \textit{ JHEP} \textbf{ 05} (2006) 026,
\href{http://www.arXiv.org/abs/hep-ph/0603175}{\texttt{ arXiv:hep-ph/0603175}}.

\bibitem{Corcella:2000bw}
\hrefCMSnoop {} {G.~Corcella {et~al.}, ``{HERWIG 6.5: an Event Generator for
  Hadron Emission Reactions With Interfering Gluons (including supersymmetric
  processes)}'',} \textit{ JHEP} \textbf{ 01} (2001) 010,
\href{http://www.arXiv.org/abs/hep-ph/0011363}{\texttt{ arXiv:hep-ph/0011363}}.

\bibitem{Frixione:2002ik}
\hrefCMSnoop {} {S.~Frixione and B.~R. Webber, ``{Matching NLO QCD Computations
  and Parton Shower Simulations}'',} \textit{ JHEP} \textbf{ 06} (2002) 029,
\href{http://www.arXiv.org/abs/hep-ph/0204244}{\texttt{ arXiv:hep-ph/0204244}}.

\bibitem{Frixione:2003ei}
\hrefCMSnoop {} {S.~Frixione, P.~Nason, and B.~R. Webber, ``{Matching NLO QCD
  and Parton Showers in Heavy Flavour production}'',} \textit{ JHEP} \textbf{
  08} (2003) 007,
\href{http://www.arXiv.org/abs/hep-ph/0305252}{\texttt{ arXiv:hep-ph/0305252}}.

\bibitem{Frixione:2008ym}
\hrefCMSnoop {} {S.~Frixione and B.~R. Webber, ``{The MC@NLO 3.4 Event
  Generator}'',}
\href{http://www.arXiv.org/abs/0812.0770}{\texttt{ arXiv:0812.0770}}.

\bibitem{Cacciari:2008zb}
\hrefCMSnoop {} {M.~Cacciari, S.~Frixione, M.~L. Mangano{ et~al.}, ``{Updated
  Predictions for the Total Production Cross Sections of Top and of Heavier
  Quark Pairs at the Tevatron and at the LHC}'',} \textit{ JHEP} \textbf{ 09}
  (2008) 127, \href{http://www.arXiv.org/abs/0804.2800}{\texttt{
  arXiv:0804.2800}}.
\href{http://dx.doi.org/10.1088/1126-6708/2008/09/127}{\texttt{
  doi:10.1088/1126-6708/2008/09/127}}.

\bibitem{Maltoni:2002qb}
\hrefCMSnoop {} {F.~Maltoni and T.~Stelzer, ``{MadEvent: Automatic Event
  Generation with MadGraph}'',} \textit{ JHEP} \textbf{ 02} (2003) 027,
\href{http://www.arXiv.org/abs/hep-ph/0208156}{\texttt{ arXiv:hep-ph/0208156}}.

\bibitem{Alwall:2007st}
\hrefCMSnoop {} {J.~Alwall {et~al.}, ``{MadGraph/MadEvent v4: The New Web
  Generation}'',} \textit{ JHEP} \textbf{ 09} (2007) 028,
\href{http://www.arXiv.org/abs/0706.2334}{\texttt{ arXiv:0706.2334}}.

\bibitem{Jung:2000hk}
\hrefCMSnoop {} {H.~Jung and G.~P. Salam, ``{Hadronic Final State Predictions
  from CCFM: The hadron- level Monte Carlo Generator CASCADE}'',} \textit{ Eur.
  Phys. J.} \textbf{ C19} (2001) 351,
  \href{http://www.arXiv.org/abs/hep-ph/0012143}{\texttt{
  arXiv:hep-ph/0012143}}.
\href{http://dx.doi.org/10.1007/s100520100604}{\texttt{
  doi:10.1007/s100520100604}}.

\bibitem{Catani:1990eg}
\hrefCMSnoop {} {S.~Catani, M.~Ciafaloni, and F.~Hautmann, ``{High-energy
  factorization and small x heavy flavor production}'',} \textit{ Nucl. Phys.}
  \textbf{ B366} (1991) 135.
\href{http://dx.doi.org/10.1016/0550-3213(91)90055-3}{\texttt{
  doi:10.1016/0550-3213(91)90055-3}}.

\bibitem{Field:2010su}
R.~Field, ``{Studying the Underlying Event at CDF and the LHC}'', in \textit{
  {Proceedings of the First International Workshop on Multiple Partonic
  Interactions at the LHC MPI'08, October 27-31, 2008}}, P.~Bartalini and
  L.~Fan{\'o}, eds.
\newblock Perugia, Italy, October, 2009.
\newblock \href{http://www.arXiv.org/abs/1003.4220}{\texttt{ arXiv:1003.4220}}.

\bibitem{Pumplin:2002vw}
\hrefCMSnoop {} {J.~Pumplin {et~al.}, ``{New Generation of Parton Distributions
  with Uncertainties from Global QCD Analysis}'',} \textit{ JHEP} \textbf{ 07}
  (2002) 012,
\href{http://www.arXiv.org/abs/hep-ph/0201195}{\texttt{ arXiv:hep-ph/0201195}}.

\bibitem{Agostinelli:2002hh}
\hrefCMSnoop {} {{ GEANT4} Collaboration, ``{GEANT4: A Simulation Toolkit}'',}
  \textit{ Nucl. Instrum. Meth.} \textbf{ A506} (2003) 250.
\href{http://dx.doi.org/10.1016/S0168-9002(03)01368-8}{\texttt{
  doi:10.1016/S0168-9002(03)01368-8}}.

\bibitem{Alwall:2008qv}
\hrefCMSnoop {} {J.~Alwall, S.~de~Visscher, and F.~Maltoni, ``{QCD Radiation in
  the Production of Heavy Colored Particles at the LHC}'',} \textit{ JHEP}
  \textbf{ 02} (2009) 017, \href{http://www.arXiv.org/abs/0810.5350}{\texttt{
  arXiv:0810.5350}}.
\href{http://dx.doi.org/10.1088/1126-6708/2009/02/017}{\texttt{
  doi:10.1088/1126-6708/2009/02/017}}.

\bibitem{Jung:2010si}
\hrefCMSnoop {} {H.~Jung {et~al.}, ``{The CCFM Monte Carlo generator CASCADE
  2.2.0}'',} \textit{ Eur. Phys. J.} \textbf{ C70} (2010) 1237,
  \href{http://www.arXiv.org/abs/1008.0152}{\texttt{ arXiv:1008.0152}}.
\href{http://dx.doi.org/10.1140/epjc/s10052-010-1507-z}{\texttt{
  doi:10.1140/epjc/s10052-010-1507-z}}.

\bibitem{Deak:2009xt}
\hrefCMSnoop {} {M.~Deak, F.~Hautmann, H.~Jung{ et~al.}, ``{Forward Jet
  Production at the Large Hadron Collider}'',} \textit{ JHEP} \textbf{ 09}
  (2009) 121.

\bibitem{Cacciari:2008gp}
\hrefCMSnoop {} {M.~Cacciari, G.~P. Salam, and G.~Soyez, ``{The anti-$k_t$ Jet
  Clustering Algorithm}'',} \textit{ JHEP} \textbf{ 04} (2008) 063,
  \href{http://www.arXiv.org/abs/0802.1189}{\texttt{ arXiv:0802.1189}}.
\href{http://dx.doi.org/10.1088/1126-6708/2008/04/063}{\texttt{
  doi:10.1088/1126-6708/2008/04/063}}.

\bibitem{CMS-PAS-TRK-10-005}
\hrefCMSnoop {} {{ CMS} Collaboration, ``{Tracking and Primary Vertex Results
  in First 7 TeV Collisions}'',} \textit{ CMS-Note} \textbf{
  CERN-CMS-PAS-TRK-10-005} (2010).

\bibitem{Khachatryan:2010pw}
\hrefCMSnoop {} {{ CMS} Collaboration, ``{CMS Tracking Performance Results from
  early LHC Operation}'',} \textit{ Eur. Phys. J.} \textbf{ C70} (2010) 1165,
  \href{http://www.arXiv.org/abs/1007.1988}{\texttt{ arXiv:1007.1988}}.
\href{http://dx.doi.org/10.1140/epjc/s10052-010-1491-3}{\texttt{
  doi:10.1140/epjc/s10052-010-1491-3}}.

\bibitem{CMS-PAS-PFT-10-002}
\hrefCMSnoop {} {{ CMS} Collaboration, ``Commissioning of the Particle-Flow
  Reconstruction in Minimum-Bias and Jet Events from pp Collisions at 7~TeV'',}
  \textit{ CMS Note} \textbf{ CERN-CMS-PAS-PFT-10-002} (2010).

\bibitem{CMS-PAS-JME-10-010}
\hrefCMSnoop {} {{ CMS} Collaboration, ``{Jet Energy Corrections Determination
  at $\sqrt{s}=7\,$ TeV}'',} \textit{ CMS Note} \textbf{
  CERN-CMS-PAS-JME-10-010} (2010).

\bibitem{CMS-PAS-JME-10-003}
\hrefCMSnoop {} {{ CMS} Collaboration, ``{Jet Performance in pp Collisions at
  $\sqrt{s}=7\,$ TeV}'',} \textit{ CMS Note} \textbf{ CERN-CMS-PAS-JME-10-003}
  (2010).

\bibitem{Fruhwirth:2007hz}
\hrefCMSnoop {} {R.~Fr{\"u}hwirth, W.~Waltenberger, and P.~Vanlaer, ``{Adaptive
  Vertex Fitting}'',} \textit{ J. Phys.} \textbf{ G34} (2007) N343.
\href{http://dx.doi.org/10.1088/0954-3899/34/12/N01}{\texttt{
  doi:10.1088/0954-3899/34/12/N01}}.

\bibitem{ref:AVR}
\hrefCMSnoop {} {W.~Waltenberger, ``{Adaptive Vertex Reconstruction}'',}
  \textit{ CMS Note} \textbf{ CERN-CMS-NOTE-2008-033} (2008).

\bibitem{Khachatryan:2010ez}
\hrefCMSnoop {} {{ CMS} Collaboration, ``{First Measurement of the Cross
  Section for Top-Quark Pair Production in Proton-Proton Collisions at
  sqrt(s)=7 TeV}'',} \textit{ Phys. Lett.} \textbf{ B695} (2011) 424,
  \href{http://www.arXiv.org/abs/1010.5994}{\texttt{ arXiv:1010.5994}}.
\href{http://dx.doi.org/10.1016/j.physletb.2010.11.058}{\texttt{
  doi:10.1016/j.physletb.2010.11.058}}.

\bibitem{Aad:2010ey}
\hrefCMSnoop {} {{ Atlas} Collaboration, ``{Measurement of the top quark-pair
  production cross section with ATLAS in pp collisions at $\sqrt{s}=7\TeV$}'',}
\href{http://www.arXiv.org/abs/1012.1792}{\texttt{ arXiv:1012.1792}}.

\end{thebibliography}\endgroup

\cleardoublepage\appendix\section{The CMS Collaboration \label{app:collab}}\begin{sloppypar}\hyphenpenalty=5000\widowpenalty=500\clubpenalty=5000\textbf{Yerevan Physics Institute,  Yerevan,  Armenia}\\*[0pt]
V.~Khachatryan, A.M.~Sirunyan, A.~Tumasyan
\vskip\cmsinstskip
\textbf{Institut f\"{u}r Hochenergiephysik der OeAW,  Wien,  Austria}\\*[0pt]
W.~Adam, T.~Bergauer, M.~Dragicevic, J.~Er\"{o}, C.~Fabjan, M.~Friedl, R.~Fr\"{u}hwirth, V.M.~Ghete, J.~Hammer\cmsAuthorMark{1}, S.~H\"{a}nsel, C.~Hartl, M.~Hoch, N.~H\"{o}rmann, J.~Hrubec, M.~Jeitler, G.~Kasieczka, W.~Kiesenhofer, M.~Krammer, D.~Liko, I.~Mikulec, M.~Pernicka, H.~Rohringer, R.~Sch\"{o}fbeck, J.~Strauss, A.~Taurok, F.~Teischinger, P.~Wagner, W.~Waltenberger, G.~Walzel, E.~Widl, C.-E.~Wulz
% ...
\vskip\cmsinstskip
% ...
1:~~Also at CERN, European Organization for Nuclear Research, Geneva, Switzerland\\
% ...

\end{sloppypar}
\end{document}